\documentclass[a4paper,12pt]{article}
\usepackage[utf8]{inputenc}
\usepackage[english]{babel}
\usepackage{epsfig}
\usepackage{multirow}
\usepackage{wrapfig}
\usepackage{subfigure}
\usepackage{amssymb,amsmath,array}
\usepackage{float}
\newcommand{\reff}[1]{(\ref{#1})}

\title{On QCD analysis of stucture function $F_2^{\gamma}$ in alternative approach}
\author{Jiri Hejbal\\
Institute of Physics of the Academy of Sciences, Czech Republic}

\begin{document}

\maketitle
\begin{abstract}
The alternative approach to QCD analysis of the photon structure function
$F_2^{\gamma}$ is presented. It differs from the conventional one by the
presence of the terms which in conventional approach appear in higher
orders. We show that this difference concerns also the photonic parton distribution functions. In the alternative approach, the complete LO analysis of $F_2^{\gamma}$ can be performed as all required quantities are known. At the NLO, however, one of the coefficient function is so far not available and thus only the photonic parton distribution function can be computed and compared to those of standard approach. We discuss the numerical difference of these approaches at the LO and the NLO approximation and show that in case of $F_2^{\gamma}$ this difference is non-negligible and may play an important role in the analysis on photon data of the future experiments.
\end{abstract}

\newpage

\section{Introduction}

The concept of the structure of the photon has appeared more then 30 years ago and is still very important phenomenological tool for analyzing processes involving photons. It is based on the fact that photon fluctuate into various states consisting of leptons, quarks, W$^{\pm}$ bosons etc. Through the primary splitting 

\begin{equation}
\gamma \rightarrow q\bar{q}
\label{gamma->qq}
\end{equation}
followed by subsequent interactions of the $q\bar{q}$ pair, the hadronic structure of the photon in the perturbative QCD can be revealed.

This structure can be described, similarly as for the case of nucleon, in terms of parton distribution functions (PDF), satisfying DGLAP evolution equations. However, for the photon these equations differ from those of nucleon one by the presence of inhomogeneous terms which are the direct consequence of \reff{gamma->qq}. This inhomogeneity has important implications for the QCD analysis of $F_{2}^{\gamma}$. In particular, one has to be careful in the interpretation of the order of QCD approximation of physical quantities, like $F_2^\gamma$, involving PDF of the photon. The way of handling this problem in most of analyses of structure function $F_2^{\gamma}$, which we call the conventional approach, has been critically discussed in \cite{Chyla1} and the alternative one was proposed\footnote{The alternative approach was also adopted for instance in \cite{kraw_1}.}. According to \cite{Chyla1}, the main source of the confusion concerning the QCD analysis of $F_{2}^{\gamma}$ stems from the definition of what the leading order (LO) and next-to-leading order (NLO) means in the context of photonic interactions. As a result, the conventional approach mixes purely QED effects with those of QCD origin. The alternative approach, compared to the conventional one, includes at each fixed order of QCD a few additional quantities. In LO the complete analysis in this approach can be performed because all required quantities are known. In NLO all but one of these quantities are known but even the remaining can in principle be derived from the known calculations \cite{VAM}. The work on this is in progress \cite{Private}.
 
In the present article we extend the alternative approach to the case of full $F_2^\gamma$ and present detailed numerical results. In particular, we emphasize the numerical importance of all additional terms as well as the overall difference between these two approaches and show that this difference is phenomenologically relevant and comparable with errors of the existing data. Finally we present the global analysis of $F_{2}^{\gamma}$ based on the model of ref. \cite{kraw}.

The paper is organized as follows. In the next Section the basic facts concerning PDF of the photon are recalled, followed in Section 3 by discussion of the QCD analysis of the structure function $F_2^{\gamma}$ at the LO and NLO. In Section 4 numerical comparison of the conventional and the alternative approaches to the QCD analysis of $F_2^{\gamma}$ at the LO is performed, followed in Section 5 by results of the global fits in $FFNS_{CJKL}$ model and the conclusion in Section 6.

\newpage

\section{Photonic parton distributions and their evolution}

Similarly as for nucleous, we can introduce the so-called "dressed" photonic PDFs, which result from the resummation of multiparton collinear emission from the corresponding "bare" parton distribution. As a result, photonic PDFs acquire dependence on the factorization scale M. This dependence is determined by the system of coupled inhomogeneous evolution equations

\begin{eqnarray}
\frac{\textup{d} \Sigma^{\gamma}(x,M)}{\textup{d}\ln M^2} &=&  \delta_{\Sigma}k_q+ P_{qq}\otimes \Sigma^{\gamma}+P_{qG}\otimes G^{\gamma},\nonumber\\
\frac{\textup{d} G^{\gamma}(x,M)}{\textup{d}\ln M^2} &=&  k_G+ P_{Gq}\otimes \Sigma^{\gamma}+P_{GG}\otimes G^{\gamma},\\
\frac{\textup{d} q^{\gamma}_{NS}(x,M)}{\textup{d}\ln M^2} &=& \delta_{NS}k_q+ P_{qq}\otimes q^{\gamma}_{NS} \nonumber
\label{DGLAP}
\end{eqnarray}
where
\begin{equation}
 \delta_{NS}=6n_f\left(\left\langle e^4\right\rangle -\left\langle e^2\right\rangle ^2 \right), \quad \delta_{\Sigma}=6n_f \left\langle e^2\right\rangle 
\end{equation}
and the inhomogeneous terms $k_q$ a $k_G$ result from \reff{gamma->qq} and QCD corrections to it. 

The non-singlet and the singlet distribution functions are defined as:
\begin{eqnarray}
\Sigma^\gamma(x,M)=\sum_f[q^\gamma_f(x,M)+\bar{q}^\gamma_f(x,M)],\nonumber\\
q^\gamma_{NS}(x,M)=\sum_f(e_i^2-\langle e^2 \rangle )(q^\gamma_f(x,M)+\bar{q}^\gamma_f(x,M))\nonumber
\end{eqnarray}
where \emph{f} runs over all active massless quark flavors.

To order $\alpha$ the photon-parton splitting functions $k(x,M)$ and the purely hadronic splitting functions $P(x,M)$ in Eq. $\reff{DGLAP}$ are given as power expansions in $\alpha_s(M)$

\begin{eqnarray}
k_{q}(x,M) &=& \frac{\alpha}{2 \pi} \left[ k_{q}^{(0)}(x)+\frac{\alpha_{s}(M)}{2 \pi}k_{q}^{(1)}(x)+\left( \frac{\alpha_{s}(M)}{2 \pi} \right)^{2} k_{q}^{(2)}(x)+ \cdots \right], \label{k(x)}\\
k_{G}(x,M) &=& \frac{\alpha}{2 \pi} \left[ \frac{\alpha_{s}(M)}{2 \pi}k_{G}^{(1)}(x)+\left( \frac{\alpha_{s}(M)}{2 \pi} \right)^{2} k_{G}^{(2)}(x)+ \cdots \right], \\
P_{ij}(x,M) &=& \frac{\alpha_{s}(M)}{2 \pi} P_{ij}^{(0)}(x) + \left( \frac{\alpha_{s}(M)}{2 \pi} \right)^{2} P_{ij}^{(1)}(x)+ \cdots \label{P(x)}
\end{eqnarray}
where the splitting functions $k_q^{(i)}$ for $i=0,1,2$ are given in Mellin moments in \appendixname{ A}. The coupling constant $\alpha_s$ satisfies the RG equation:
\begin{equation}
\frac{\textup{d}\alpha_s(\mu)}{\textup{d}\textup{ln}\:\mu^2}= \beta(\alpha_s(\mu))\equiv-\frac{\beta_0}{4\pi}\alpha_s^2(\mu)-\frac{\beta_1}{16\pi^2}\alpha_s^3(\mu)+\cdots
\label{RGE}
\end{equation}
with $\beta_0=11-2f/3$ and $\beta_1=102-38f/3$. 
To the NLO, the solution of the equation \reff{RGE} can be approximated in the form: 
\begin{equation}
\frac{\alpha_s(\mu)}{4\pi} \doteq \frac{1}{\beta_0 \ln \mu^2 /\Lambda^2}-\frac{\beta_1}{\beta_0^3}\frac{\ln \ln \mu^2/\Lambda^2}{(\ln \mu^2/\Lambda^2)^2}.
\label{res_RGE}
\end{equation}
Equation \reff{RGE} has infinite number of solutions which can be parameterized by the parameter $\Lambda$. The choice of $\Lambda$ then specifies a particular solution of \reff{RGE}.

\begin{figure}[ht]
\centerline{\includegraphics[width=0.9\columnwidth]{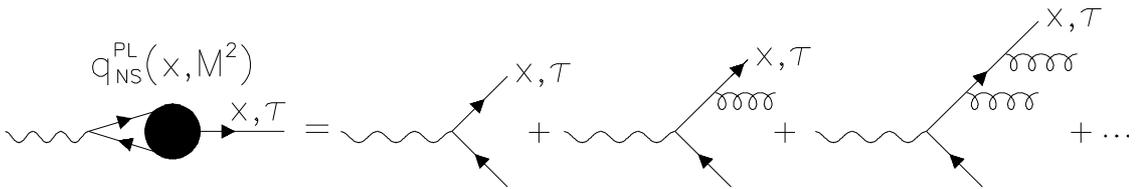}}
\caption{Diagrams defining the pointlike parts of non-singlet quark distribution function.}\label{qpointlike}
\end{figure}

The general solution of the evolution equations \reff{DGLAP} can be written as the sum

\begin{equation}
f^{\gamma}(x,M)=f^{\gamma,had}(x,M)+f^{\gamma,PL}(x,M)
\label{PL&HAD}
\end{equation}
of general solution of the corresponding homogeneous equations, called hadronic, and particular solution of the full inhomogeneous equations for which we can take the so-called pointlike solution which results from the resumation of contributions to the pure QED coupling $\gamma\rightarrow q\bar{q}$ where one gluon, two gluons etc. are emitted. In the non-singlet case the diagrams like those in Fig. \ref{qpointlike} are ressummed:

\begin{eqnarray}
\lefteqn{q_{NS}^{\gamma,PL}(x,M_{0},M) \equiv} \nonumber\\
& & \equiv \frac{\alpha}{2 \pi} k_{NS}^{(0)}(x) \int_{M_{0}^{2}}^{M^{2}} \frac{d \tau}{\tau} + \int_{x}^{1} \frac{dy}{y} P_{qq}^{(0)} \left( \frac{x}{y} \right) \int_{M_{0}^{2}}^{M^{2}} \frac{d \tau_{1}}{\tau_{1}} \frac{\alpha_{s}(\tau_{1})}{2 \pi} \frac{\alpha}{2 \pi} k_{NS}^{(0)}(y) \int_{M_{0}^{2}}^{\tau_{1}} \frac{d \tau_{2}}{\tau_{2}}+ \nonumber\\
& & \int_{x}^{1} \frac{dy}{y} P_{qq}^{(0)} \left( \frac{x}{y} \right)  \int_{y}^{1} \frac{dw}{w} P_{qq}^{(0)} \left( \frac{y}{w} \right) \int_{M_{0}^{2}}^{M^{2}} \frac{d \tau_{1}}{\tau_{1}} \frac{\alpha_{s}(\tau_{1})}{2 \pi} \times \nonumber\\
& & \times \int_{M_{0}^{2}}^{\tau_{1}} \frac{d \tau_{2}}{\tau_{2}} \frac{\alpha_{s}(\tau_{2})}{2 \pi} \frac{\alpha}{2 \pi} k_{NS}^{(0)}(w) \int_{M_{0}^{2}}^{\tau_{2}} \frac{d \tau_{3}}{\tau_{3}}+ \cdots.
\label{PL_LO_rada}
\end{eqnarray}

The evolution equations \reff{DGLAP} can not be solved analytically in $x$-space and one has to use numerical methods. One of them is based on the Mellin transform:
 
\begin{equation}
 f(n,M) \equiv \int_0^1 \textup{d}xx^{n-1}f(x,M),
\end{equation}
which converts the equations \reff{DGLAP} into much simpler ones:
\begin{eqnarray}
\frac{\textup{d} \Sigma^{\gamma}(n,M)}{\textup{d}\ln M^2} &=&  \delta_{\Sigma}k_q(n,M)+ P_{qq}(n,M) \Sigma^{\gamma}(n,M)+P_{qG}(n,M)G^{\gamma}(n,M),\label{PDFn_1}\\
\frac{\textup{d} G^{\gamma}(n,M)}{\textup{d}\ln M^2} &=&  k_G(n,M)+ P_{Gq}(n,M) \Sigma^{\gamma}(n,M)+P_{GG}(n,M) G^{\gamma}(n,M),\label{PDFn_2}\\
\frac{\textup{d} q^{\gamma}_{NS}(n,M)}{\textup{d}\ln M^2} &=& \delta_{NS}k_q(n,M)+ P_{qq}(n,M) q^{\gamma}_{NS}(n,M).
\label{PDFn_3}
\end{eqnarray}
Evolution equations \reff{PDFn_1}-\reff{PDFn_3} are solvable analytically. In the non-singlet case, taking into account only $k^{(0)}$ and $P^{(0)}$ we get \cite{GR} :

\begin{eqnarray}
q^{\gamma,PL}_{NS}(n,M) &\!\!\!\!=\!\!\!\!& \frac{4 \pi} {\alpha_s(M)} \left(1-\left(\frac{\alpha_s(M)}{\alpha_s(M_0)}\right)^{1-2P^{(0)}_{qq}(n)/\beta_0}\right)a \label{PL_LO},\!\! \quad a\!=\!\frac{1}{1-\frac{2}{\beta_0}P^{(0)}_{qq}(n)}\frac{\alpha}{2\pi\beta_0}k^{(0)}(n),\nonumber\\
q^{\gamma,had}_{NS}(n,M) &\!\!\!\!=\!\!\!\!&\left(\frac{\alpha_s(M)}{\alpha_s(M_0)}\right)^{-2P^{(0)}_{qq}(n)/\beta_0}q^{\gamma,had}(n,M_0)
\label{HAD_LO}
\end{eqnarray}
Adding also $k^{(1)}$ and $P^{(1)}$ we find:
\begin{eqnarray}
\!\!\!\!\!\!\!\!\!\!&\!\!\!\!\!\!\!\!\!\!&q^{\gamma,PL}_{NS}(n,M)= \frac{4\pi}{\alpha_s(M)}\left[1+\frac{\alpha_s(M)}{2\pi}U_{NS} \right]\left[1-\left(\frac{\alpha_s(M)}{\alpha_s(M_0)} \right)^{1-2P_{qq}^{(0)}(n)/\beta_{0}} \right]a \nonumber\\
\!\!\!\!\!\!\!\!\!\!&\!\!\!\!\!\!\!\!\!\!&+\left[1-\left(\frac{\alpha_s(M)}{\alpha_s(M_0)} \right)^{-2P_{qq}^{(0)}(n)/\beta_{0}} \right]\frac{1}{-P_{qq}^{(0)}(n)}\frac{\alpha}{2\pi}\left[k^{(1)}(n)-\frac{\beta_1}{2\beta_0}k^{(0)}(n)-U_{NS}k^{(0)}(n) \right], \label{PL_NLO}\\
\!\!\!\!\!\!\!\!\!\!&\!\!\!\!\!\!\!\!\!\!&q^{\gamma,had}_{NS}(n,M)= \left\lbrace 1+\left(\frac{\alpha_s(M)}{2 \pi}-\frac{\alpha_s(M_0)}{2 \pi}\right)U_{NS}\right\rbrace \left(\frac{\alpha_s(M)}{\alpha_s(M_0)}\right)^{-2P^{(0)}_{qq}(n)/\beta_0}q^{\gamma,had}(n,M_0) \nonumber
\label{HAD_NLO}
\end{eqnarray}
where $\alpha_s$ is now given by equation \reff{RGE}, where we keep terms proportional to $\beta_0$ and $\beta_1$ and

\begin{eqnarray}
U_{NS}&\!\!\!\!=\!\!\!\!&-\frac{2}{\beta_0}\left( P_{qq}^{(1)}-\frac{\beta_1}{2\beta_0}P_{qq}^{(0)}\right).
\end{eqnarray}
In the flavor singlet case the above formulae become more complicated because $P^{(0)}$, $P^{(1)}$, and consequently also $R$ are the 2x2 matrices. For more details see \appendixname{ B}.

Next step is to transform the distribution function from the Mellin moments back to the $x$-space. This inverse Mellin transformation can be done by integration in the complex plane:

\begin{equation}
f(x)=\frac{1}{2 \pi i} \int_{\xi -i \infty}^{\xi +i \infty} F(n) x^{-n} dn
\label{Inverse_Mellin_1}
\end{equation}
where the integration path lies to the right of the right-most singularity of F(n). 

If the function $f(x)$ is real the above formula may be simplified as:

\begin{equation}
f(x)=\frac{1}{\pi} \int_{0}^{\infty} F(\xi + i \eta) x^{-\xi - i \eta} \; d \eta.
\end{equation}
To solve DGLAP evolution equations requires the specification of the boundary conditions which contain information on x-dependence of the photonic PDFs at some arbitrary initial scale $Q_0^2$. This boundary conditions are usually parameterized by a set of parameters which are determined by the comparison with experimental data. Another parameter which have to be chosen is a real parameter $\xi$. As is seen from \reff{Inverse_Mellin_1}, it determines the position of the integral path on real axis. In fact, such a choice is arbitrary if $\xi$ lies right to the most-right singularity and the integral path runs upwards to infinity, but this is no more true if we solve this integral numerically. In such a case the optimal shape and position together with the upper limit of the integration have to be chosen in order to satisfy the requirements of the accurancy or/and computation time.

\section{QCD analysis of the structure function $F_2^{\gamma}$}
\subsection{Conventional approach at the LO and NLO}

With the explicit solutions for the photonic parton distributions $q_{NS}^{\gamma}$, $\Sigma^{\gamma}$ and $G^{\gamma}$ at hand it is now straightforward to obtain the photon structure function $F_2^{\gamma}(x,Q^2)$ given as (in the rest of the text we set $M=Q$)
\begin{eqnarray}
 \frac{1}{x}F_2^{\gamma}(x,Q^2)&=&q_{NS}^{\gamma}(Q) \otimes C_q(1)+\frac{\alpha}{2\pi}\delta_{NS}C_{\gamma}(x,1)+ \nonumber\\
 &&\left\langle e^2\right\rangle \Sigma^{\gamma}(Q) \otimes C_q(1)+\left\langle e^2\right\rangle G^{\gamma}(Q) \otimes C_G(1)+\frac{\alpha}{2\pi}\left\langle e^2\right\rangle \delta_{\Sigma}C_{\gamma}(x,1)
\label{F2gamma}
\end{eqnarray}
where the coefficient functions $C_q(x)$, $C_G(x)$ and $C_{\gamma}$ entering the $F_2^{\gamma}$ in convolution with the photon distributions admit the following perturbative expansions
\begin{eqnarray}
 C_q(x,Q/M)=& \delta(1-x)+&\frac{\alpha_{s}(Q^2)}{2\pi}C_q^{(1)}(x,Q/M)+\cdots, \label{expansion_Cq}\\
 C_G(x,Q/M)=& &\frac{\alpha_{s}(Q^2)}{2\pi}C_G^{(1)}(x,Q/M)+\cdots,\\
 C_{\gamma}(x,Q/M)=& C_{\gamma}^{(0)}(x,Q/M)+&\frac{\alpha_{s}(Q^2)}{2\pi}C_{\gamma}^{(1)}(x,Q/M)+\cdots. \label{expansion_Cgamma}
\end{eqnarray}
The coefficient functions $C_q^{(1)}$, $C_{\gamma}^{(0)}$ and $C_{\gamma}^{(1)}$ can be found in \appendixname{ A}. In order to compute the structure function $F_2^{\gamma}$ to the desired order of QCD one has to provide photonic PDFs and coefficient functions to corresponding order of perturbative QCD.

We find that in the pointlike solution \reff{PL_LO} the QCD coupling constant $\alpha_s$ appears in the denominator. If we take this fact literally i.e. as meaning that the lowest order of pQCD pointlike parts of the photonic distribution functions behave like ($1/\alpha_s$), we conclude that:
\begin{itemize}
\item at the LO the structure function $F_2^{\gamma}$ is related to the distribution functions in the same way as in the hadronic case: 

\begin{equation}
F_2^{\gamma,LO}(x,Q^2)=q_{NS}^{\gamma}(x,Q)+\left\langle e^2\right\rangle \Sigma^{\gamma}(x,Q)
\label{F2_LO_con}
\end{equation}
where non-singlet and singlet distribution functions satisfy evolution equations including terms $k_q^{(0)}$ and $P^{(0)}$ only

\item whereas at the NLO

\begin{eqnarray}
F_2^{\gamma,NLO}\!\!\!\!\!&\!\!\!\!\!&(x,Q^2)= \nonumber\\
\!\!\!\!\!&\!\!\!\!\!&\left[1+\frac{\alpha_s}{2\pi}C_{q}(1) \right]\otimes q_{NS}^{\gamma}(Q)+\nonumber \\
\!\!\!\!\!&\!\!\!\!\!& \langle e^2 \rangle \left\lbrace  \left[1+\frac{\alpha_s}{2\pi}C_{q}^{(1)}(1) \right]\otimes \Sigma^{\gamma}(Q)+\frac{\alpha_s}{2 \pi}C_{G}^{(1)}(1) \otimes G^{\gamma}(Q)\right\rbrace +\frac{\alpha}{2\pi}C_{\gamma}^{(0)}(x,1) \nonumber
\label{F2_NLO_con}
\end{eqnarray}
where non-singlet, singlet and gluon distribution functions, satisfy now the evolution equations including also the splitting functions $P^{(1)}$, $k_q^{(1)}$ and $k_G^{(1)}$.
\end{itemize}
Let us, however, ask the question what happens to the pointlike solution \reff{PL_LO} and \reff{PL_NLO} (and consequently to $F_2^{\gamma}$) if we switch off QCD by sending $\Lambda\rightarrow 0$. For this purpose, it is sufficient consider the non-singlet pointlike distribution function. It is obvious that provided $M_0$ is kept fixed when $\Lambda\rightarrow 0$ the sum \reff{PL_LO_rada} approaches its first term:

\begin{equation}
q_{NS}^{\gamma,PL}\rightarrow \frac{\alpha}{2\pi}k_{NS}^{(0)}\ln\frac{M^2}{M_0^2} \label{q_NS_QED}
\end{equation}
corresponding to the first diagram in Fig. \ref{qpointlike} and describing purely QED splitting $\gamma \rightarrow q\bar{q}$. This is obvious also from the fact that switching off QCD, keeping $M_0$ fixed, we discard in Fig. \ref{qpointlike} all diagrams with QCD verticies and are then left with the first one, purely QED one, giving \reff{q_NS_QED}. This shows that the claim  $q_{NS}^{PL} \propto 1/\alpha_s$ is misleading. In fact, this claim can be used only as a shorthand for the specification of large $Q^2$ behaviour of $q_{NS}^{\gamma,PL}$. Morover, we see that in the conventional LO approximation to $F_2^{\gamma}$ \reff{F2_LO_con}, the effects of QED are partially ignored as the splitting function $k_q^{(0)}$ is included, whereas $C_{\gamma}^{(0)}$ is not, despite the fact that both come from the first diagram in Fig. \ref{qpointlike}.

\subsection{Alternative approach}

The alternative approach \cite{Chyla1} is based on the procedure of separating the QED contributions and defining terms "leading" and "next-to-leading" etc. uniquely as terms of the QCD origin, starting at $\alpha_s$.

The crucial role in this approach plays the consideration that $q^{PL}$ is of the order $O(\alpha)$ rather then $O(1/\alpha_s)$. Consequently $q^{\gamma,PL}$ can be written as the sum of two terms:

\begin{equation}
q^{\gamma,PL}(x,M,M_0)=q_{QED}^{\gamma,PL}(x,M,M_0)+q_{QCD}^{\gamma,PL}(x,M,M_0)
\end{equation}
where the purely QED contribution proportional to $\alpha$ is given as (for single quark flavour with unit electric charge)

\begin{equation}
q_{QED}^{\gamma,PL}(x,M,M_0)=\frac{\alpha}{2\pi}3\left(x^2+(1-x)^2\right)\ln\frac{M^2}{M_0^2}
\end{equation}
where $M_0$ is a free parameter, on which the separation of hadron and pointlike parts in \reff{PL&HAD} depends, and $q_{QCD}^{\gamma}$, describing genuine QCD effects satisfies the inhomogeneous evolution equation\footnote{In the following the dependence on $M_0$ will be supressed}

\begin{eqnarray}
\frac{\textup{d}q_{QCD}^{\gamma,PL}(x,M)}{\ln M^2}&\!\!\!\!\!\!\!=\!\!\!\!\!\!\!&\frac{\textup{d}q^{\gamma,PL}(x,M)}{\ln M^2}-\frac{\textup{d}q_{QED}^{\gamma,PL}(x,M)}{\ln M^2}= \nonumber\\
&&\frac{\alpha_s}{2\pi}\left[\frac{\alpha}{2\pi}k^{(1)}+P^{(0)}\otimes q^{\gamma,PL}_{QED} \right] +\left(\frac{\alpha_s}{2\pi}\right)^{2}\left[\frac{\alpha}{2\pi}k^{(2)}+P^{(1)}\otimes q^{\gamma,PL}_{QED} \right]+\cdots \nonumber \\
&&\frac{\alpha_s}{2\pi}P^{(0)}\otimes q^{\gamma,PL}_{QCD}+\left(\frac{\alpha_s}{2\pi}\right)^{2}P^{(1)}\otimes q^{\gamma,PL}_{QCD}+\cdots.
\end{eqnarray}
Consequently, the structure function $F_2^{\gamma,PL}$ at the LO approximation includes the QED terms:

\begin{equation}
F_{2, QED}^{\gamma,PL,LO}(x,Q^2)=q_{QED}^{\gamma,PL}(x,Q)+\frac{\alpha}{2\pi}C_{\gamma}^{(0)}(x,1)
\end{equation}
and the QCD contributions proportional to $\alpha\alpha_s$:

\begin{equation}
F_{2, QCD}^{\gamma,PL,LO}(x,Q^2)=q_{QCD}^{\gamma,PL}(x,Q)+\frac{\alpha_s}{2\pi}C_q^{(1)}(1)\otimes q_{QED}^{\gamma,PL}(Q)+\frac{\alpha}{2\pi}\frac{\alpha_s}{2\pi}C_{\gamma}^{(1)}(x,1).
\label{F2_LO_QCD}
\end{equation}
By adding also hadronic part of the distribution functions and summing over charges of all active quarks and antiquarks, this leads\footnote{Note that it is consistent to use in \reff{F2_LO_alter} the full pointlike $q^{\gamma,PL}$ instead of $q_{QED}^{\gamma,PL}$} to full $F_2^{\gamma}$ at the LO approximation:
\begin{eqnarray}
\!\!\!\!\!\!\!\!\!\!\!\!\!\!\!\!\!\!&& \frac{1}{x}F_{2,alter.}^{\gamma,LO}(x,Q^2)=\nonumber \\
\!\!\!\!\!\!\!\!\!\!\!\!\!\!\!\!\!\!&&q_{NS}^{\gamma}(x,Q)+\left\langle e^2 \right\rangle  \Sigma^{\gamma}(x,Q)+\frac{\alpha_s}{2\pi}C_{q}^{(1)}(1)\otimes q^{\gamma}_{NS}(Q)+\left\langle e^2\right\rangle\frac{\alpha_s}{2\pi}C_{q}^{(1)}(1)\otimes \Sigma^{\gamma}(Q)+ \nonumber \\
\!\!\!\!\!\!\!\!\!\!\!\!\!\!\!\!\!\!&&\delta_{NS}\frac{\alpha}{2\pi}C_{\gamma}^{(0)}(x,1)+\left\langle e^2\right\rangle\delta_{\Sigma}\frac{\alpha}{2\pi}C_{\gamma}^{(0)}(x,1)+\delta_{NS}\frac{\alpha}{2\pi}\frac{\alpha_s}{2\pi}C_{\gamma}^{(x)}(x,1)+\left\langle e^2\right\rangle\delta_{\Sigma}\frac{\alpha}{2\pi}\frac{\alpha_s}{2\pi}C_{\gamma}^{(1)}(x,1).
\label{F2_LO_alter}
\end{eqnarray}
The expression \reff{F2_LO_alter} thus differs from the expression in conventional approach \reff{F2_LO_con} by

\begin{itemize}
\item 
the appearance of the corresponding contributions of the photonic coefficient functions $C_{\gamma}^{(0)}$ and $C_{\gamma}^{(1)}$,
\item
the appearance of the convolution of quark coefficient function $C_q^{(1)}$ with $q_{NS}^{\gamma}$ and $\Sigma^{\gamma}$,
\item
the fact that $k_{NS}^{(1)}$ resp. $k_q^{(1)}$ and $k_G^{(1)}$ are included in the evolution equations for $q_{NS}^{\gamma}$ resp. $\Sigma^{\gamma}$.
\end{itemize}

\subsection{Alternative approach at the NLO}

At the NLO we add in both the splitting and coefficient functions the terms standing by $\alpha_s^2$.

In Tab.\ref{table:1} we list all quantities used in the both alternative and conventional approaches at the LO and NLO. At present, all quantities except the coefficient function $C_{\gamma}^{(2)}$ are known. This allows us to perform the global analysis of $F_{2}^{\gamma}$ up to the NLO in the conventional approach. In the alternative approach, only LO global analysis is possible to perform. Nevertheless, we will make some comments on the calculation in the NLO in the alternative approach as well. For this purpose, we focus on the non-singlet part of $F_2^{\gamma}$.

In the standard approach the non-singlet pointlike distribution function satisfies in the NLO the following equation:

\begin{eqnarray}
\lefteqn{\frac{\textup{d} q^{\gamma}_{NS}(x,M)}{\textup{d}\ln M^2} =}\nonumber \\
&& \frac{\alpha}{2\pi} k_q^{(0)}(x)+\frac{\alpha}{2\pi} \frac{\alpha_s(M)}{2\pi} k_q^{(1)}(x)+\left(\frac{\alpha_s(M)}{2\pi}P_{qq}^{(0)}+\left(\frac{\alpha_s(M)}{2\pi}\right)^{2}P_{qq}^{(1)}\right) \otimes q^{\gamma}_{NS}(M)
\label{EQ_NLO_CON}
\end{eqnarray}
whereas in the alternative approach, we include also the term proportional to $k_q^{(2)}$

\begin{figure}[t]
  \begin{center}
    \subfigure[]{\label{figg:edge2-a}\includegraphics[scale=0.405]{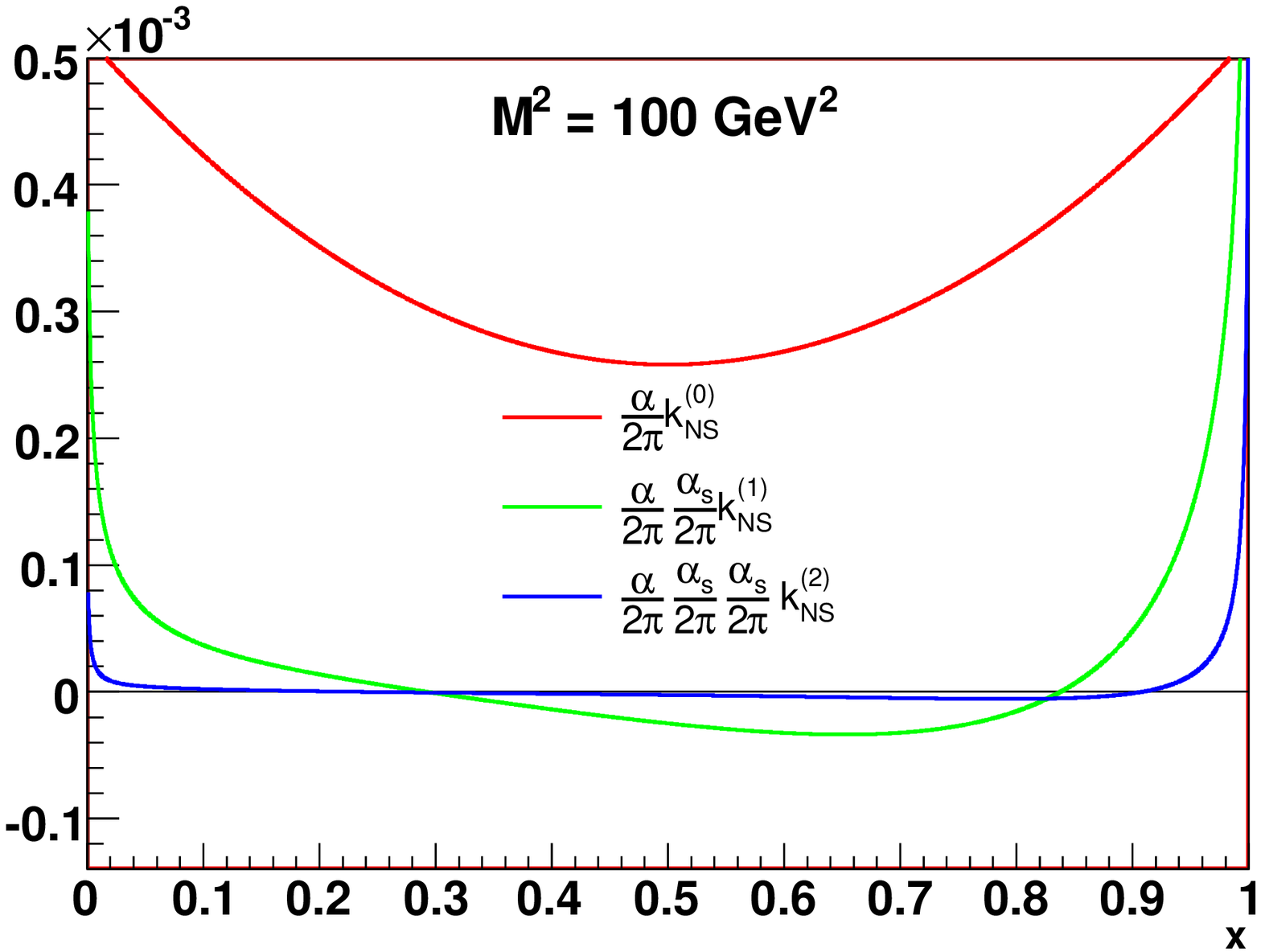}}
    \subfigure[]{\label{figg:edge2-b}\includegraphics[scale=0.405]{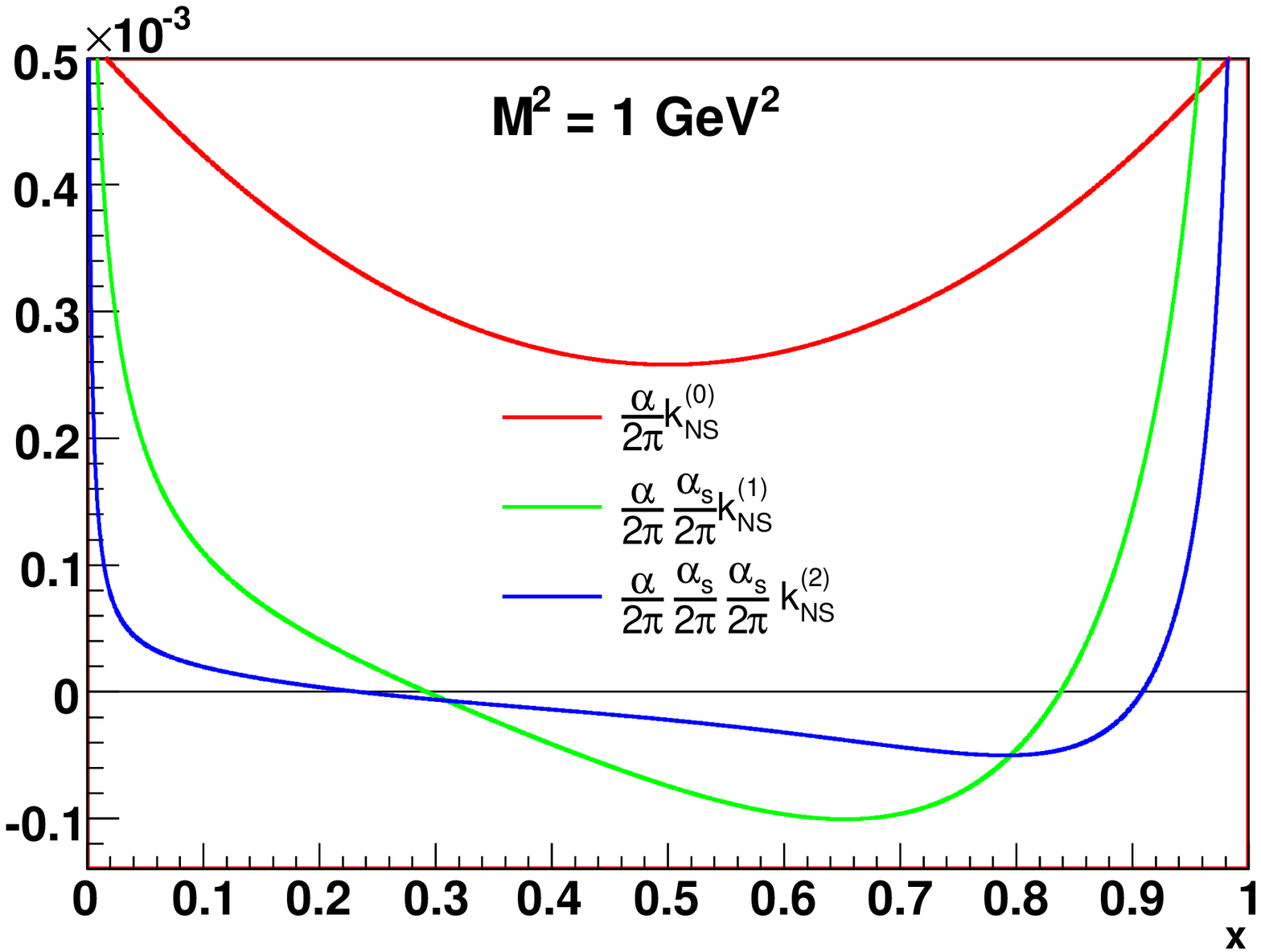}}
  \end{center}
  \caption{Comparison of the first three terms of the expansion of the photonic splitting function $k(x)$ for $M^2=100\  \textup{GeV}^2$ (a) and $M^2=1\  \textup{GeV}^2$ (b).}
  \label{k_s}
\end{figure}

\begin{table*}[t]
\newcommand{\m}{\hphantom{$-$}}
\newcommand{\cc}[1]{\multicolumn{1}{c}{#1}}
\renewcommand{\tabcolsep}{0.6pc} 
\renewcommand{\arraystretch}{1.45} 
\begin{tabular}{@{}lllll}
\hline
           & Conventional  & Alternative \\
\hline
LO QCD       & $k_q^{(0)}$, $P^{(0)}$ & + $k_q^{(1)}$, $C_{\gamma}^{(0)}$, $C_{\gamma}^{(1)}$, $C_{q}^{(1)}$\\
NLO QCD     & $k_q^{(0)}$, $P^{(0)}$, $k_q^{(1)}$, $k_G^{(1)}$, $P^{(1)}$, $C_{\gamma}^{(0)}$, $C_{q}^{(1)}$, $C_{G}^{(1)}$  & + $k_q^{(2)}$, $k_G^{(2)}$, $C_{\gamma}^{(1)}$, $C_{\gamma}^{(2)}$, $C_{q}^{(2)}$, $C_{G}^{(2)}$ \\
\hline
\end{tabular}
\caption{The list of all quantities entering to the definition of $F_2^{\gamma}$ in alternative and conventional approach in leading and next-to-leading order.}
\label{table:1}
\end{table*}

\begin{eqnarray}
\!\!\!\!\!\!\!\!\!&&\frac{\textup{d} q^{\gamma}_{NS}(x,M)}{\textup{d}\ln M^2} = \nonumber \\
\!\!\!\!\!\!\!\!\!&& \frac{\alpha}{2\pi}\left( k_q^{(0)}(x)+ \frac{\alpha_s}{2\pi} k_q^{(1)}(x)+ \left(\frac{\alpha_s}{2\pi}\right)^{2} k_q^{(2)}(x) \right) +\left(\frac{\alpha_s}{2\pi}P_{qq}^{(0)}+\left(\frac{\alpha_s}{2\pi}\right)^{2}P_{qq}^{(1)}\right) \otimes q^{\gamma}_{NS}(M).
\label{EQ_NLO_ALT}
\end{eqnarray}
We now investigate the numerical effect of such a modification on the solution of \reff{EQ_NLO_ALT}. In Fig. \ref{k_s} the comparison of three terms of the expansion of the photonic splitting function $k_q(x)$ proportional to $k_q^{(0)}$, $k_q^{(1)}$ and $k_q^{(2)}$  is shown. The relative importance of this contributions depends on M$^2$. In Fig. \ref{k_s}a and Fig. \ref{k_s}b they are calculated for M$^{2}=100$ GeV$^2$ and M$^{2}=1\ $GeV$^2$. One can expect that adding the term $k_q^{(2)}$ to the evolution equation has only small numerical effect in comparison to the corresponding inclusion of the term proportional to $k_q^{(1)}$. 

\begin{figure}[ht]
\centerline{\includegraphics[width=0.7\columnwidth]{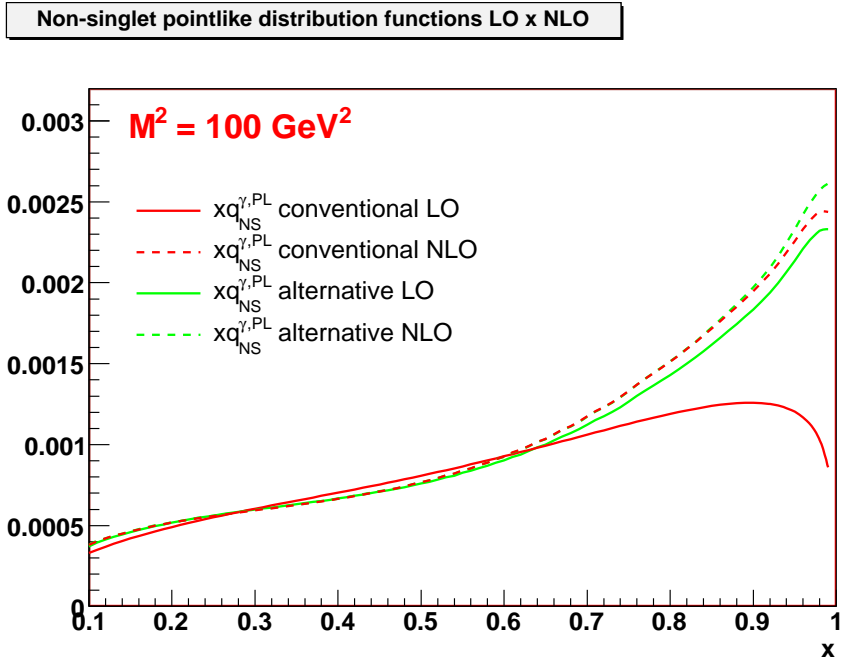}}
\caption{Comparison of the pointlike distribution functions in conventional and alternative approach at the LO and NLO for $M^2=100\ \textup{GeV}^2$ and $\Lambda=313\ \textup{MeV}$}\label{PL_LOxNLO}
\end{figure}

Solving evolutions \reff{EQ_NLO_CON} and \reff{EQ_NLO_ALT} we find that non-singlet pointlike distribution function at the NLO in the alternative approach differs from the conventional one by the term:

\begin{eqnarray}
q_{NS,alter.}^{\gamma,PL,NLO}-q_{NS,conv.}^{\gamma,PL,NLO}=\frac{\alpha_s(M)}{\left( -1- \frac{2P^{(0)}}{\beta_0} \right)}\left( {\frac {\alpha\,k_q^{(2)}}{2{\pi }^{2}\beta_0}}-{\frac {\alpha\,k_q^{(1)}\,\beta_1}{4{\pi }^{2}{\beta_0}^{2}}} \right) \left( 1- \left( {\frac {\alpha_s(M) }{\alpha_s(M_0) }} \right) ^{-1- \frac{2P^{(0)}}{\beta_0}} \right) \nonumber
\end{eqnarray}
which indeed represents only a small correction even in the region $x$ close to 1. The comparison of the non-singlet pointlike distribution functions in both approaches at the LO and NLO is shown in Fig. \ref{PL_LOxNLO}. In this plot we can see the effect of adding the term proportional to $k^{(1)}$ in the case of the LO approximation and the term proportional to $k^{(2)}$ in the case of the NLO approximation (this effect is given by the difference of the solid and doted lines of the same colour).

It is obvious that in the conventional approach the difference between the LO and NLO approximations of the parton distribution functions is much bigger than in the alternative one. In the conventional approach this difference is given by adding $P^{(1)}$ and dominantly by $k_q^{(1)}$. Let us recall that in the alternative approach $k_q^{(1)}$ enters the definition of the LO pointlike distribution. Consequently, the difference between the two approaches is much bigger at the LO than in the NLO. More precisely, the pointlike distribution function at the LO in the alternative approach is numerically close to the pointlike distribution function in the standard conventional approach in the NLO. The small difference between pointlike distribution functions in both approaches in the NLO is due to the small numerical effect of the term proportional to $k^{(2)}$ discussed above.

The non-singlet part of $F_2^{\gamma}$ in the alternative approach has the following form:
\begin{eqnarray}
\!\!\!\!\!\!\!\!\!\!\!\!\!\!\!&&\frac{1}{x}F_{2,alter.}^{\gamma,NLO}(x,Q^2)=q_{NS}^{\gamma}(x,Q)+ \frac{\alpha_s}{2\pi}C_{q}^{(1)}(1)\otimes q^{\gamma}_{NS}(Q)+\nonumber\\
\!\!\!\!\!\!\!\!\!\!\!\!\!\!\!&&\left(\frac{\alpha_s}{2\pi}\right)^{2}C_{q}^{(2)}(1)\otimes q^{\gamma}_{NS}(Q)+\delta_{NS}\frac{\alpha}{2\pi}\left(C_{\gamma}^{(0)}(x,1)+\frac{\alpha_s}{2\pi}C_{\gamma}^{(1)}(x,1)+\left(\frac{\alpha_s}{2\pi}\right)^{2}C_{\gamma}^{(2)}(x,1)\right)
\label{F2_NLO_alter}
\end{eqnarray}
which differs from the conventional definition \reff{F2_NLO_con} by the presence of $C_{\gamma}^{(1)}$, $C_{\gamma}^{(2)}$ and $C_{q}^{(1)}$, entering through the convolution with the photonic distribution functions. 

\begin{figure}[ht]
\centerline{\includegraphics[width=0.7\columnwidth]{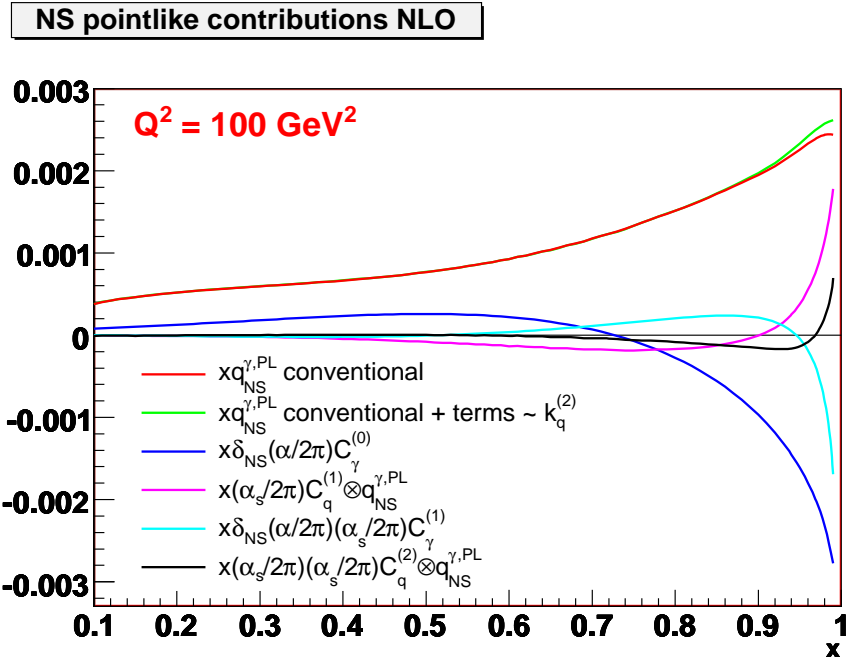}}
\caption{Contributions to NLO QCD expression for the pointlike part of $F_{2}^{\gamma}$ in the alternative approach for $Q^2=100\  \textup{GeV}^2$}\label{prispevky_F2_NLO}
\end{figure}

As already mentioned, coefficient function $C_{\gamma}^{(2)}$ has not so far been calculated. This alows us to perform in the alternative approach only LO global analysis of $F_2^{\gamma}$. To point out the difference between the conventional and alternative approach at the NLO, it is sufficient to restrict our discussion to the pointlike part of the non-singlet photonic distribution functions. The graphic representation of all available contributions entering to the formula \reff{F2_NLO_alter} is shown in Fig. \ref{prispevky_F2_NLO}.

\section{Numerical comparison of the conventional and the alternative approaches to $F_2^{\gamma}$ at the LO}
We now compare the two approaches to the QCD analysis of the non-singlet, singlet and full pointlike part of $F_2^{\gamma}$ at the LO. We concentrate on the numerical consequences of including the additional terms in the alternative approach. We discuss numerical effects of all these terms separately as well as the overall differences between these two approaches.

\begin{figure}[tb]
  \begin{center}
    \subfigure[]{\label{fig:edge-a}\includegraphics[scale=0.405]{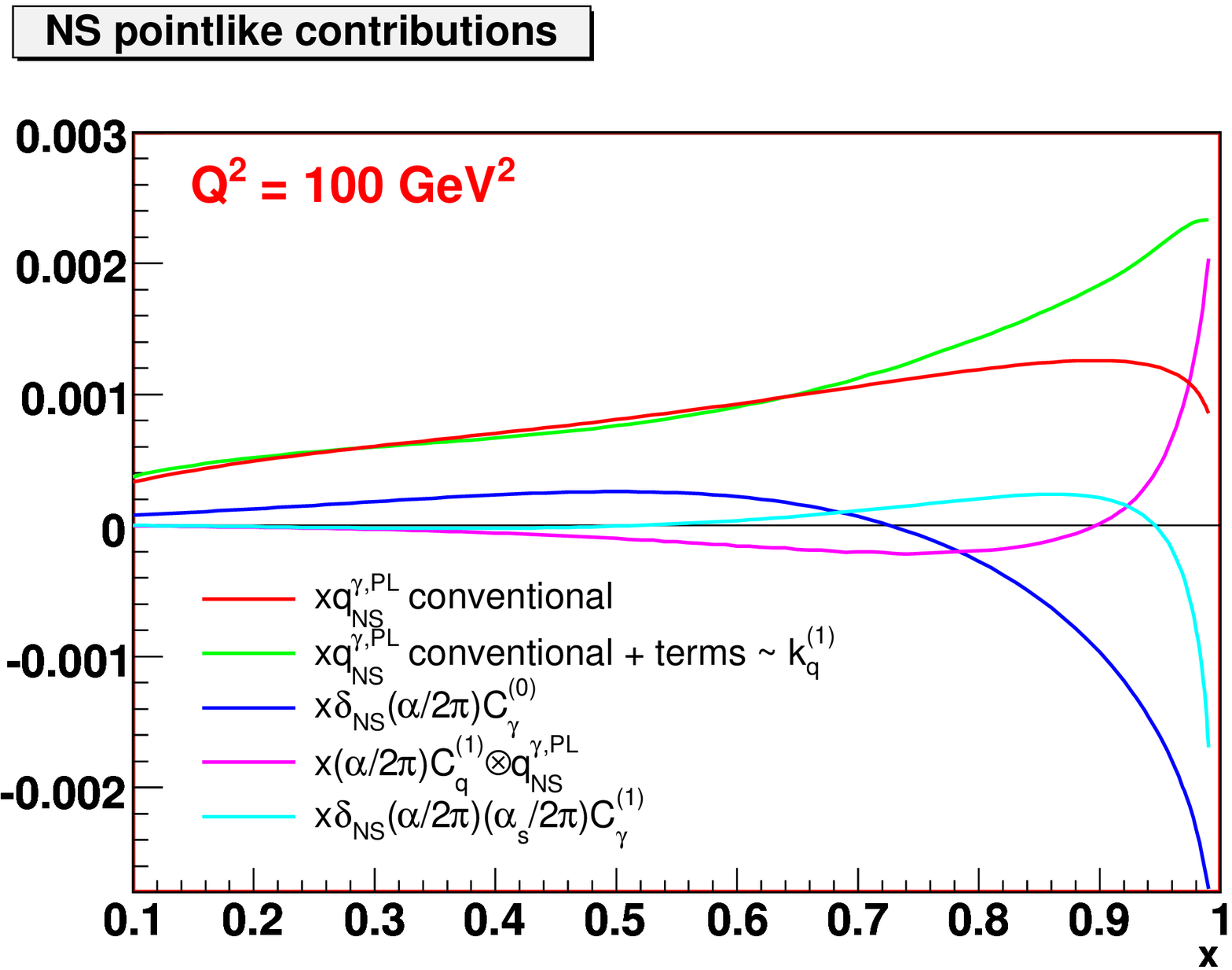}}
    \subfigure[]{\label{fig:edge-b}\includegraphics[scale=0.405]{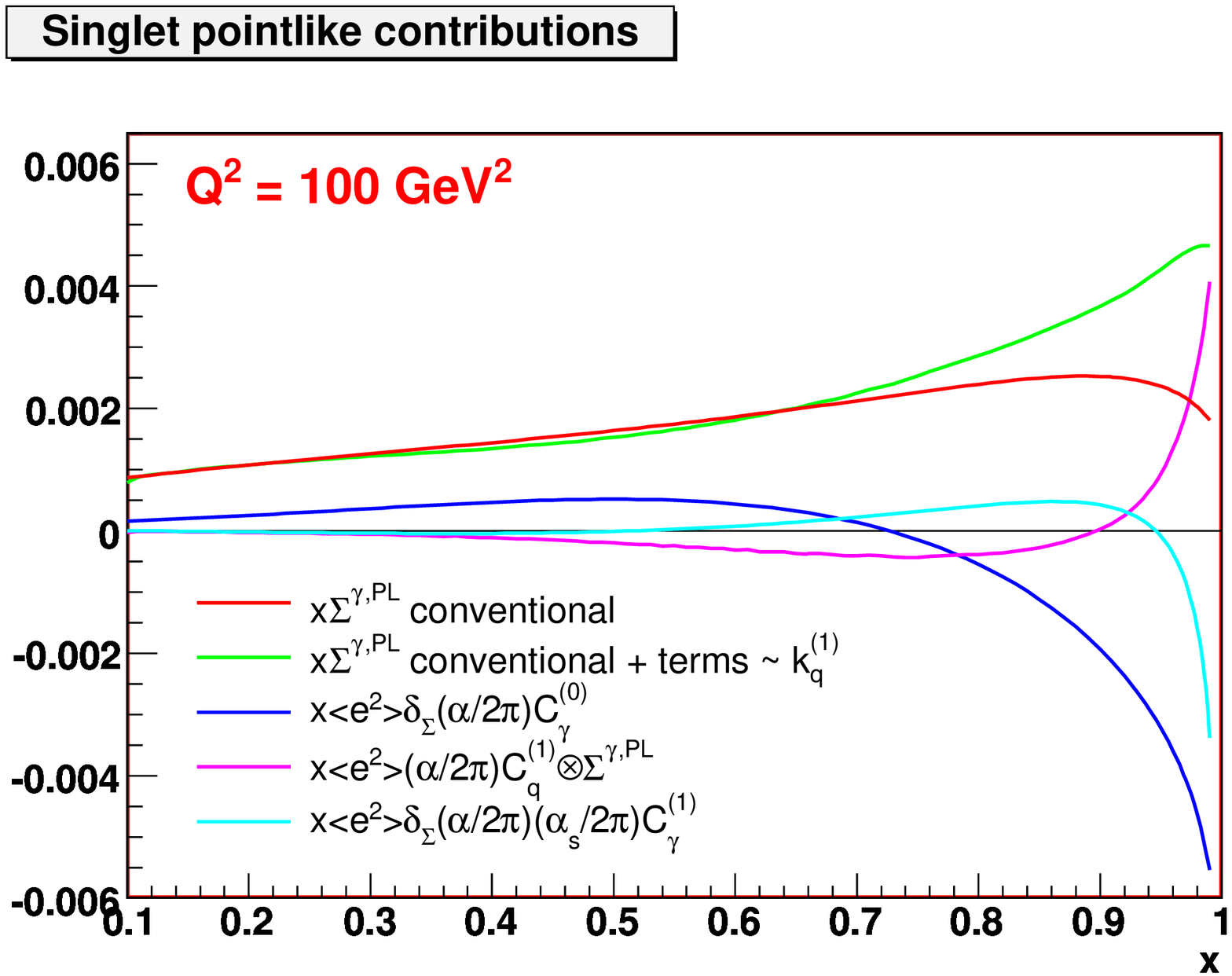}}
    \subfigure[]{\label{fig:edge-c}\includegraphics[scale=0.405]{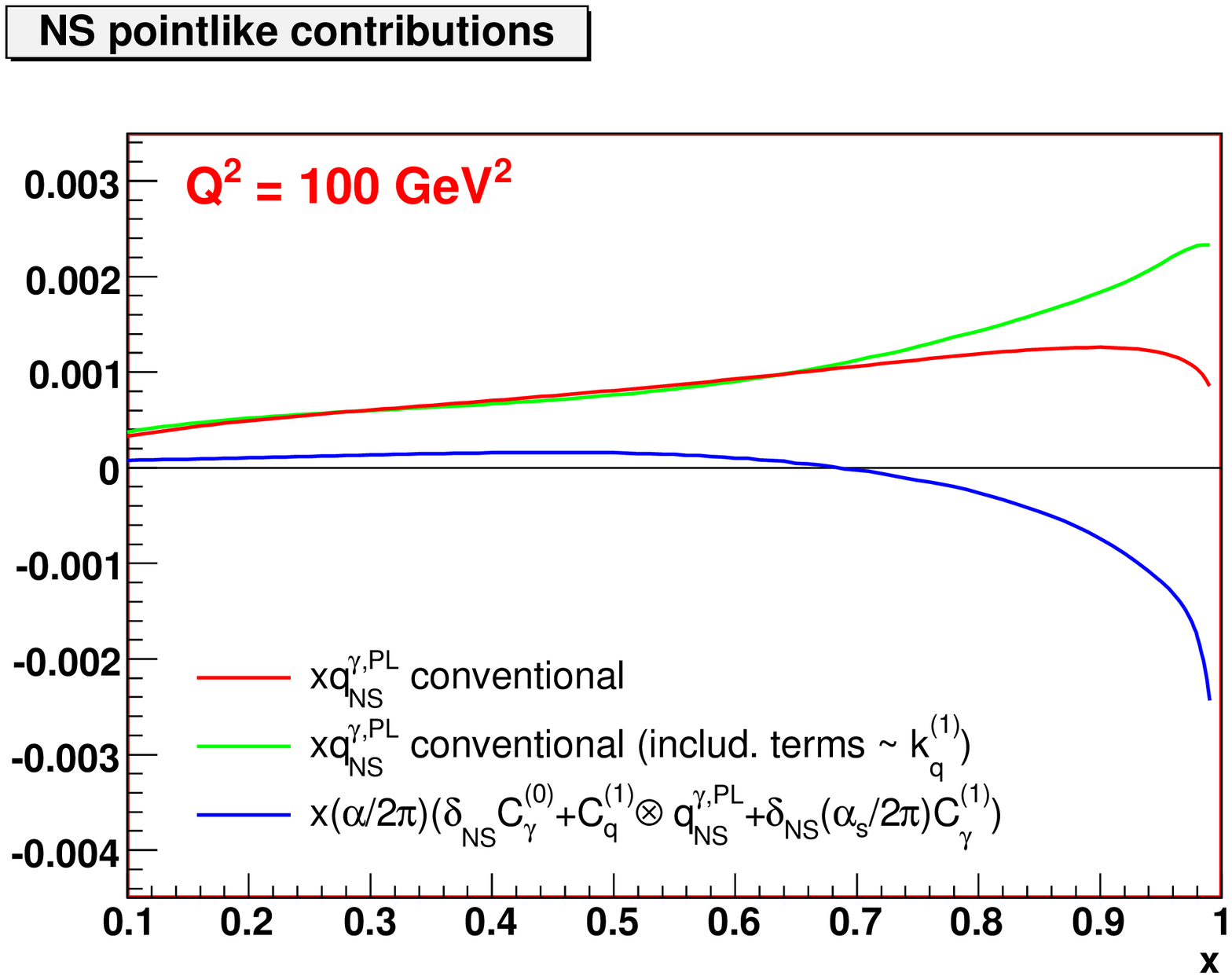}}
    \subfigure[]{\label{fig:edge-d}\includegraphics[scale=0.405]{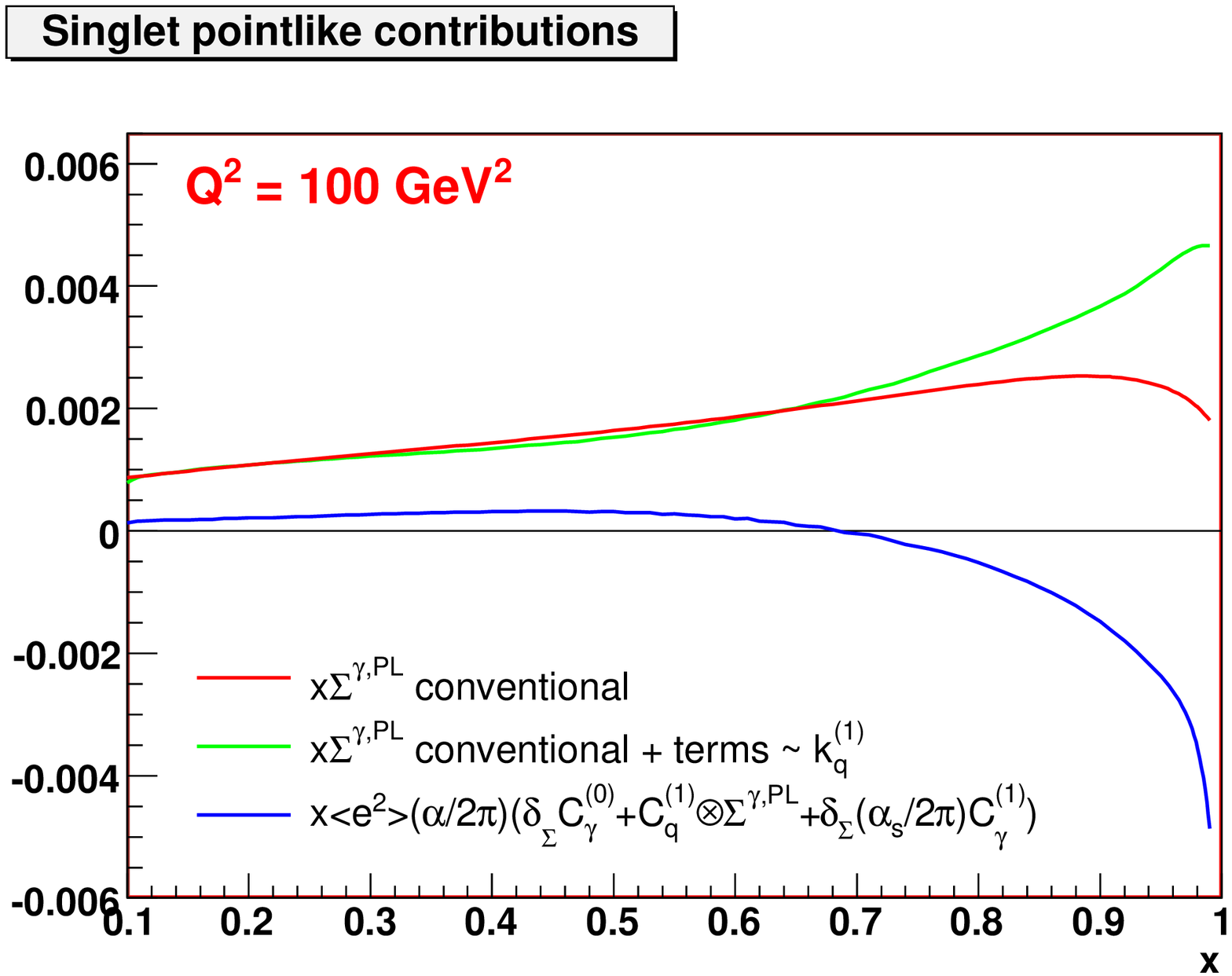}}
    \subfigure[]{\label{fig:edge-e}\includegraphics[scale=0.405]{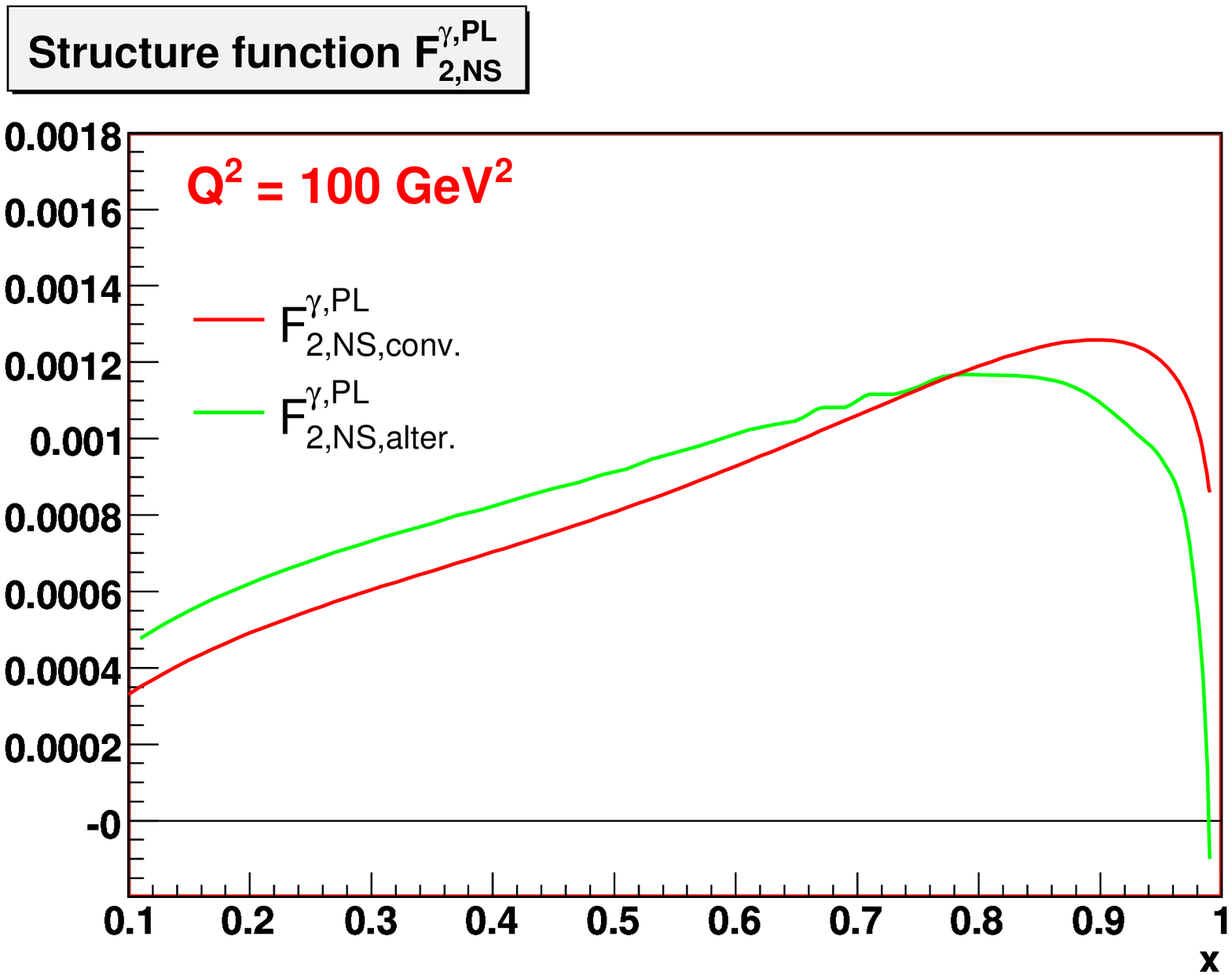}}
    \subfigure[]{\label{fig:edge-f}\includegraphics[scale=0.405]{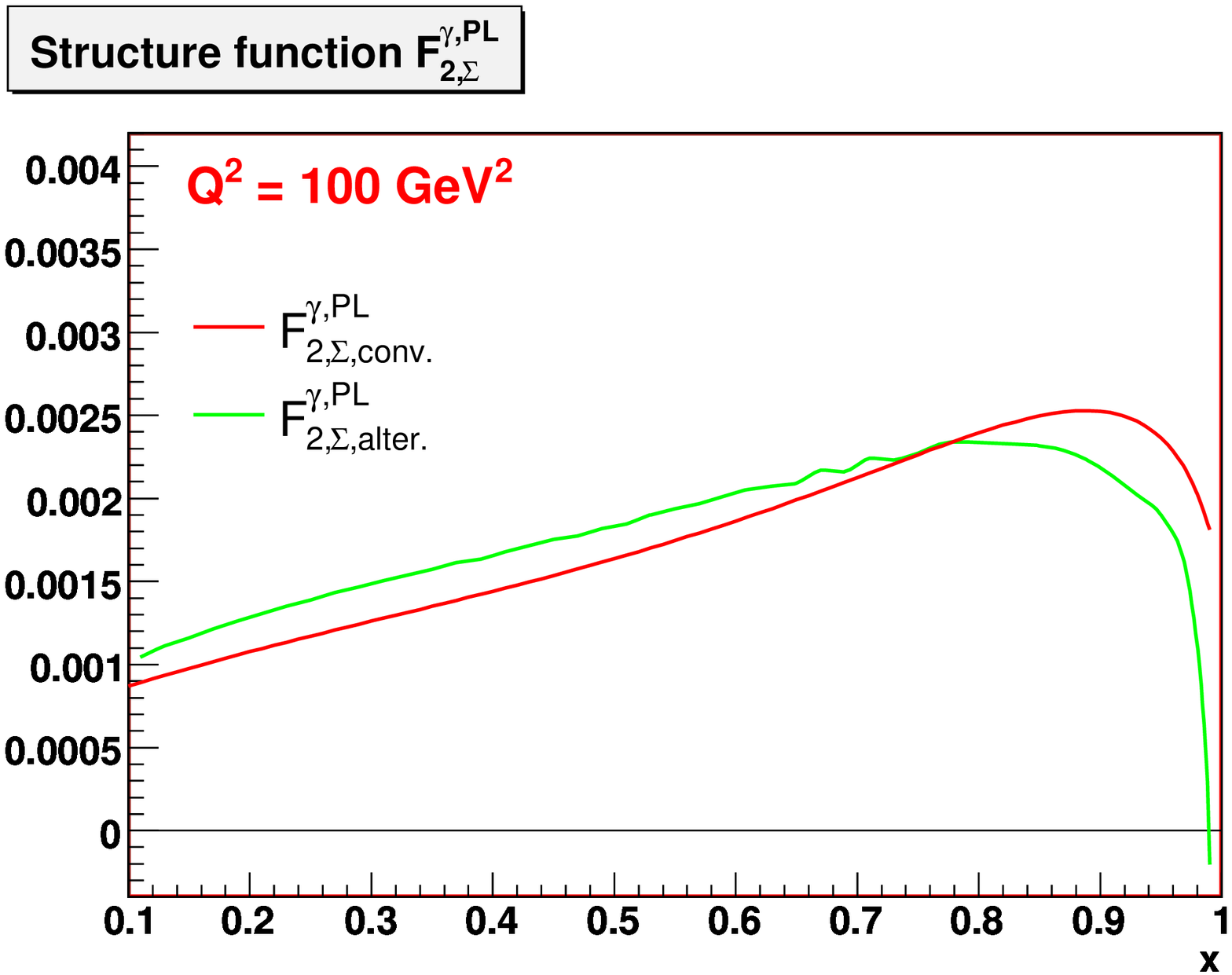}}
  \end{center}
  \caption{Contributions to the LO QCD expression for the non-singlet pointlike part of $F_{2}^{\gamma}$ (a), (c), (e) and singlet pointlike part of $F_{2}^{\gamma}$ (b), (d), (f) included in alternative approach for $Q^2=100\  \textup{GeV}^2$.}
  \label{NS&S_prispevky}
\end{figure}

The individual contributions to the non-singlet and the singlet part of $F_2^{\gamma,PL}$ in the alternative approach are plotted in Fig. \ref{NS&S_prispevky}a and Fig. \ref{NS&S_prispevky}b and compared to the non-singlet and the singlet part of $F_2^{\gamma,PL}$ in the conventional approach.

All contributions included in the alternative approach evolve with $Q^2$ (except the term proportional to $C_{\gamma}^{(0)}$). As seen from Fig. \ref{NS&S_prispevky}c and Fig. \ref{NS&S_prispevky}d, in both non-singlet and singlet cases, the sum of the terms proportional to the coefficient functions $C_{\gamma}^{(0)}$ and $C_{\gamma}^{(1)}$ give the positive contribution in the region $x \lesssim 0.7$. The region $x\gtrsim0.7$ is dominated by the negative contribution of $C_{\gamma}^{(0)}$ which is strengthened by negative contribution of $ C_{\gamma}^{(1)}$ close to $x= 1$.

The contribution of $ C_{q}^{(1)}$, entering through the convolution with distribution function $q_{NS}^{\gamma}$ and $\Sigma^{\gamma}$ give the positive correction close to $x= 1$.

Finally, adding the splitting function $k_q^{(1)}$ into the evolution equations represents an important positive contribution for their solutions in the region $x \gtrsim 0.65$. This is valid for both non-singlet and singlet distribution function.

Putting all contributions together we compare in the Fig. \ref{NS&S_prispevky}e and Fig. \ref{NS&S_prispevky}f the non-singlet and the singlet $F_{2}^{\gamma,PL}$ in the conventional and the alternative approach. In the alternative approach, $F_{2,NS}^{\gamma,PL}$ and $F_{2,\Sigma}^{\gamma,PL}$ lie higher in the region up to $x\doteq 0.8$. On the contrary, in the region close to $x= 1$, the values of $F_{2,NS}^{\gamma,PL}$ and $F_{2,\Sigma}^{\gamma,PL}$ in the alternative approach decrease much faster then in the conventional one. This is mainly due to the inclusion of $C^{(0)}_{\gamma}$.

\begin{figure}[htb]
  \begin{center}
    \subfigure[]{\label{fig:edge2-a}\includegraphics[scale=0.405]{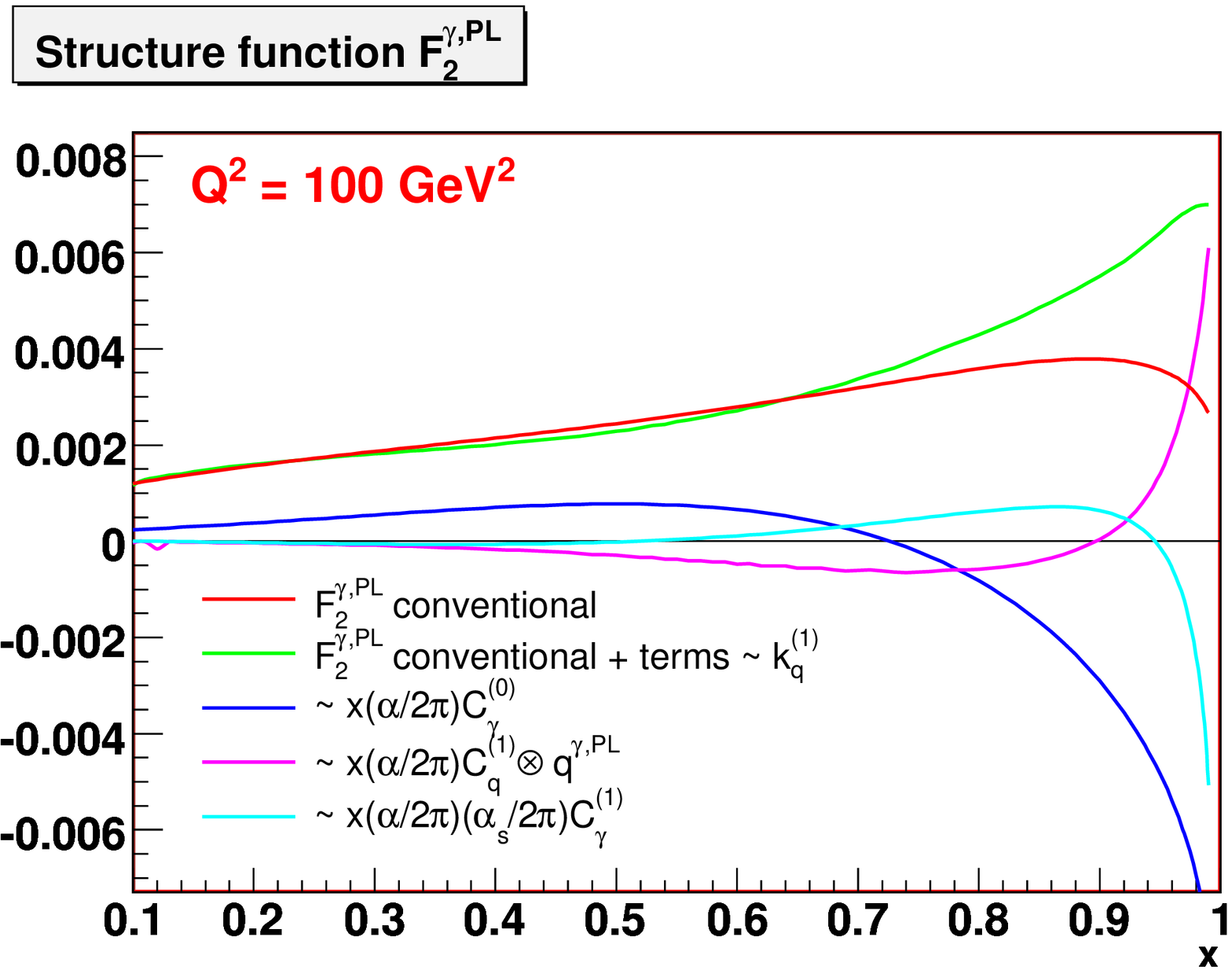}}
    \subfigure[]{\label{fig:edge2-b}\includegraphics[scale=0.40]{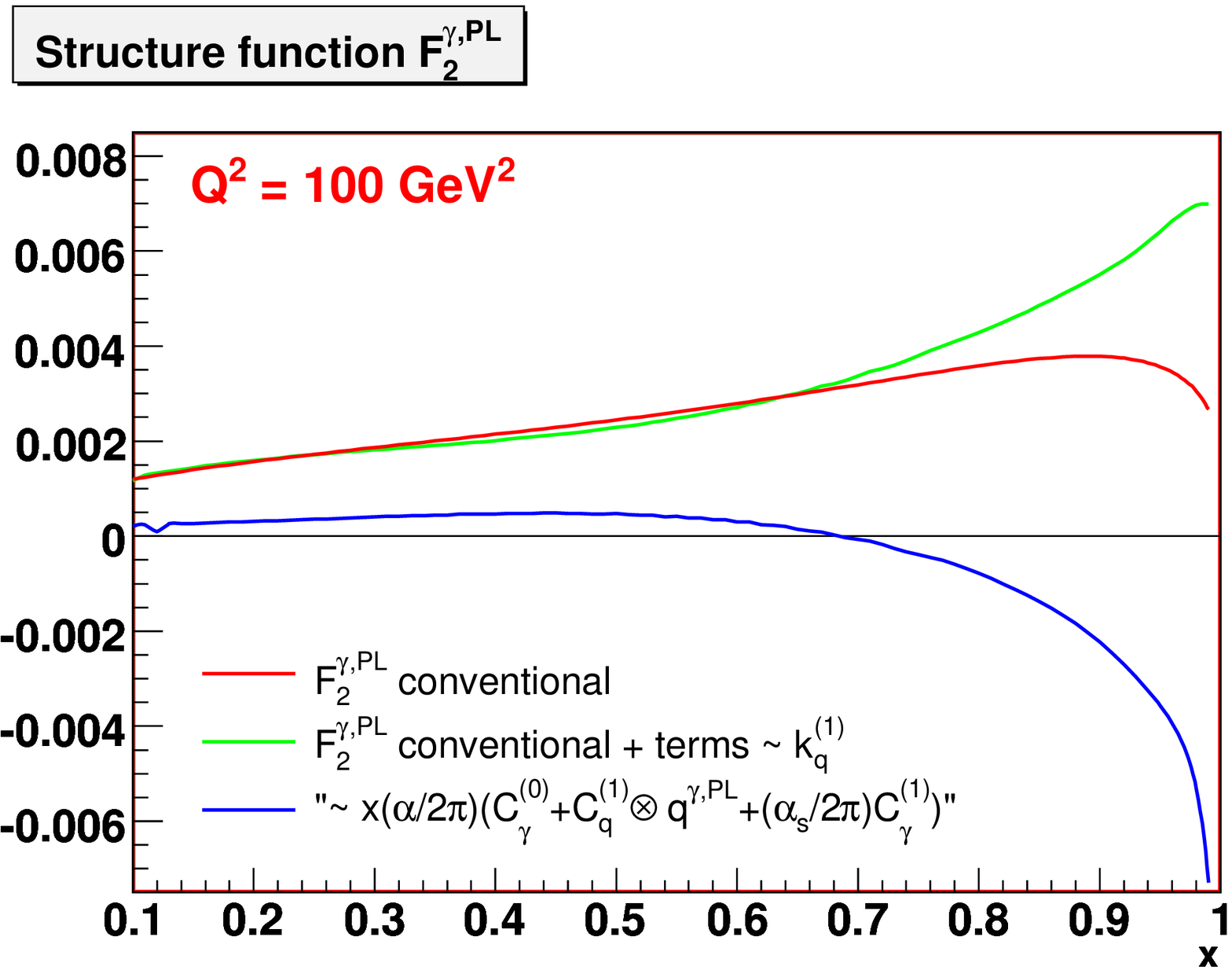}}
    \subfigure[]{\label{fig:edge2-c}\includegraphics[scale=0.40]{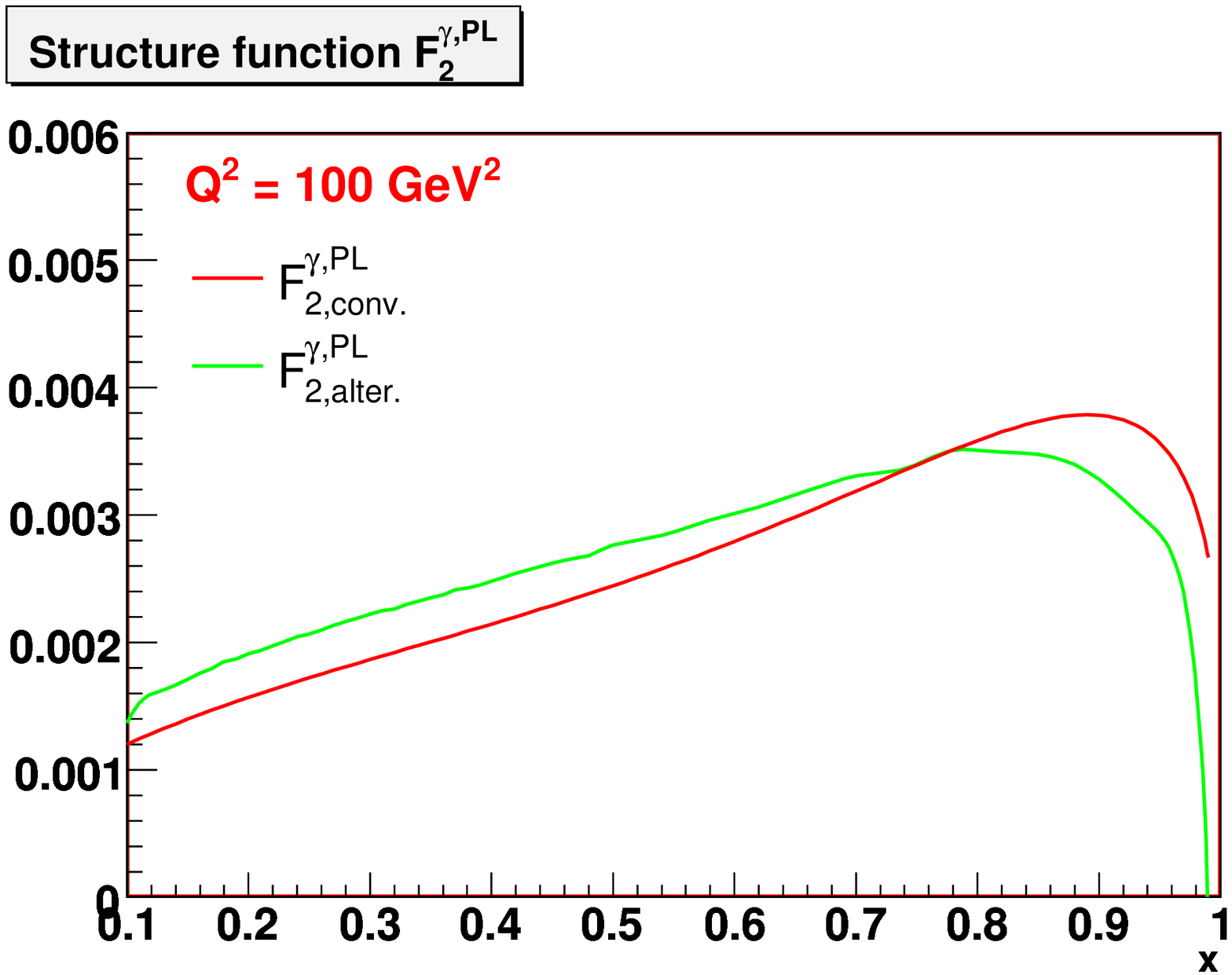}}
  \end{center}
  \caption{Comparison of the full pointlike $F_{2}^{\gamma}$ at the LO in the conventional and the alternative approach for $Q^2=100\  \textup{GeV}^2$.}
  \label{F2_PL_prispevky}
\end{figure}

In the Fig. \ref{F2_PL_prispevky} we plot the non-singlet and the singlet parts to investigate the behaviour of the full pointlike $F_2^{\gamma}$ in both approaches. In Fig. \ref{F2_PL_prispevky}a all the additional contributions are plotted separately. As seen from Fig. \ref{F2_PL_prispevky}b, the sum of terms $\sim C_{\gamma}^{(0)}$, $\sim C_{\gamma}^{(1)}$ and $\sim C_q^{(1)} \otimes q$ rapidely decrease in the region close to $x=1$. This is compensated by the term $\sim k_q^{(1)}$. Putting all terms together (Fig. \ref{F2_PL_prispevky}c) allow us to compare the full pointlike $F_2^{\gamma}$ in both approaches.

\begin{figure}[t]
  \begin{center}
    {\includegraphics[scale=0.405]{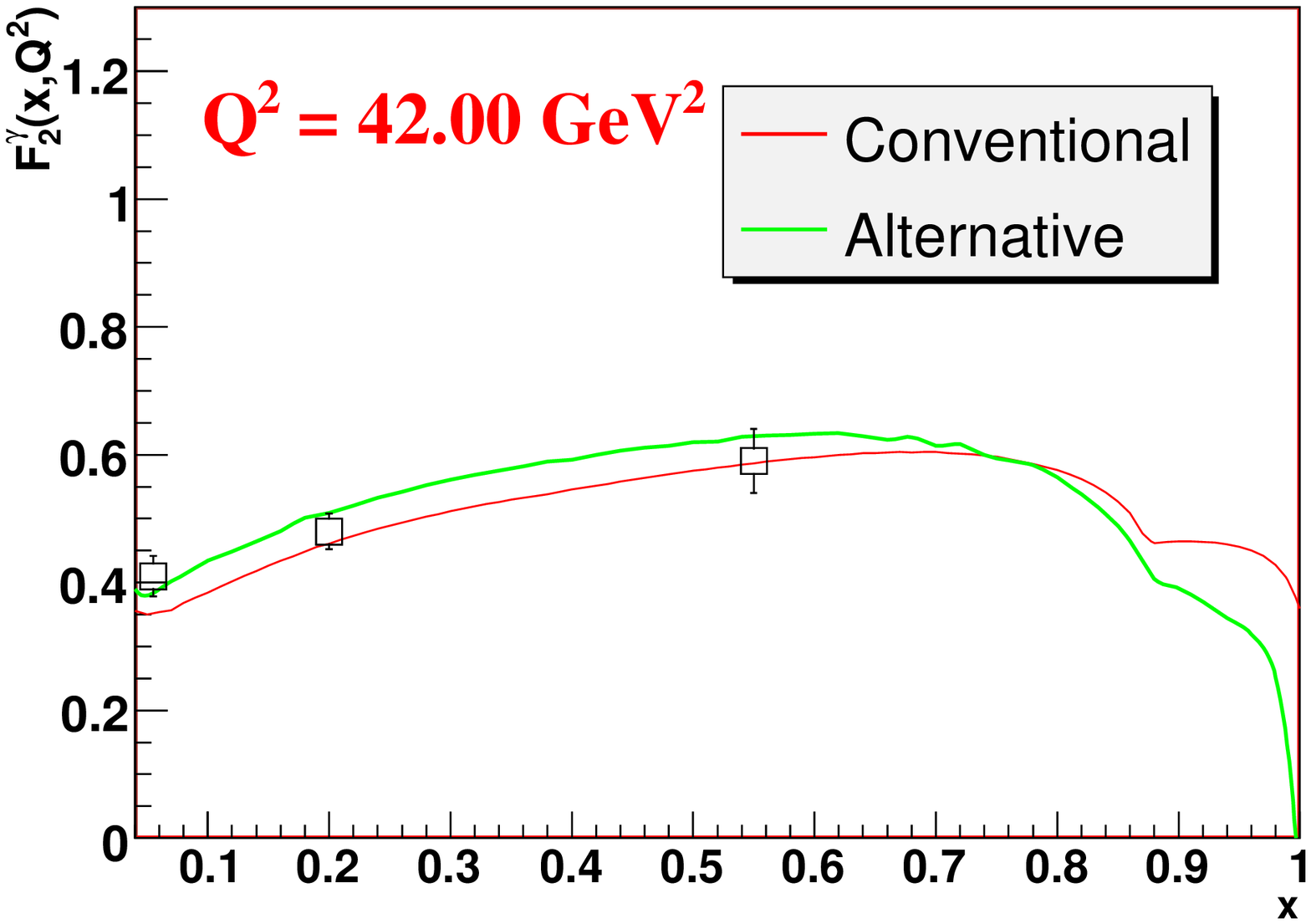}}
    {\includegraphics[scale=0.405]{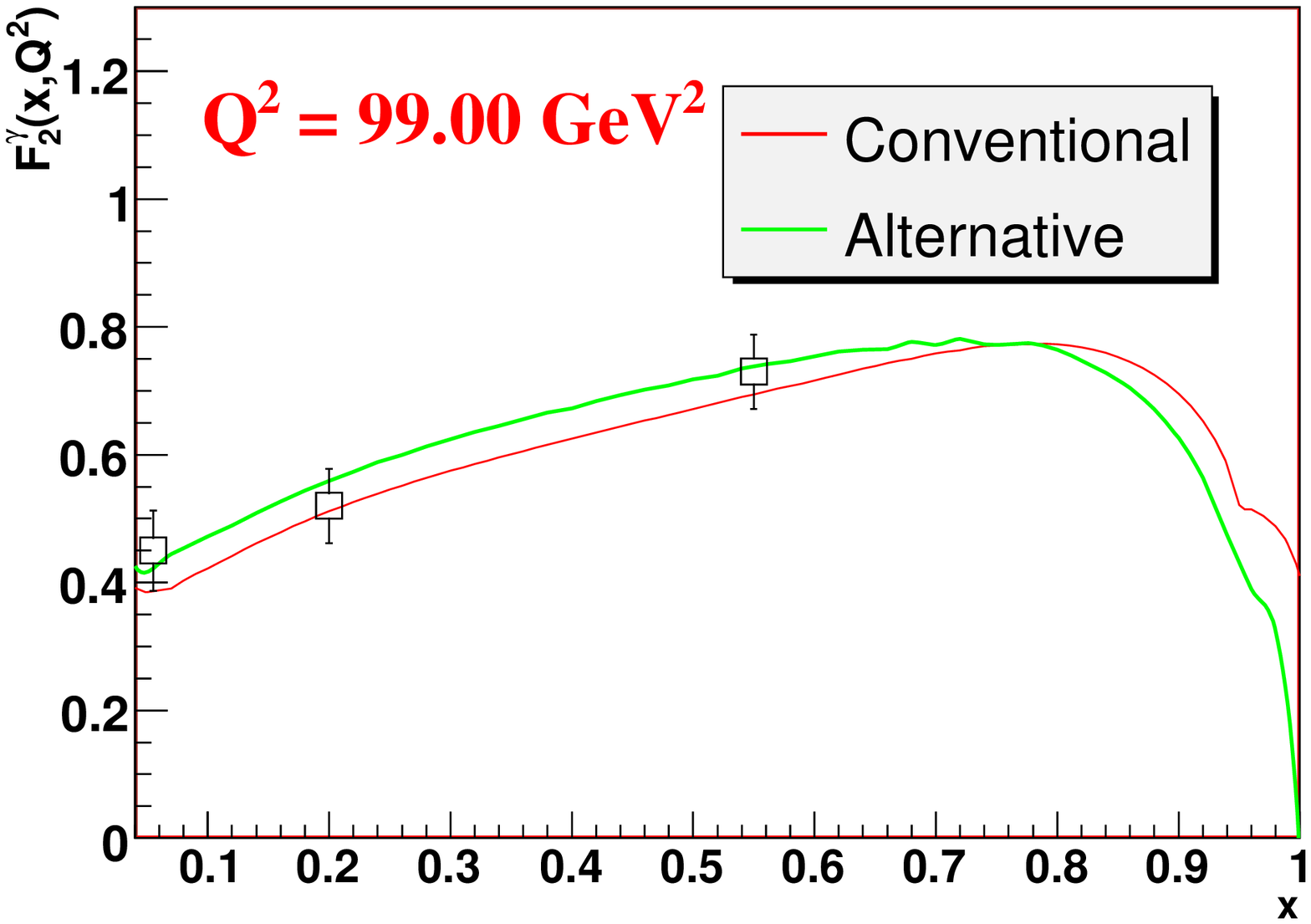}}
  \end{center}
\caption{Comparison of the experimental data of $F_2^{\gamma}$ with the solutions of the evolution equations in the conventional and alternative approach. We used parametrization of $FFNS_{CJKL}$ model with parameters of the first row of Tab. \ref{srr1}.}
\label{ts1}
\end{figure}

The significance of numerical difference between both approaches becomes clearer if we compare these differences with errors of data on $F_{2}^{\gamma}$. In the kinematical regions where the errors are reasonably small we conclude that this difference is numerically important as illustrated in Fig. \ref{ts1}. Is evident that the numerical difference between both approaches is comparable to the accuracy of measurement of $F_{2}^{\gamma}$.

\clearpage

\clearpage

\section{Fits to data of $F_2^{\gamma}$ in the alternative approach}

In this section we present an exploratory study on global analysis of the structure function $F_2^{\gamma}$ at the LO in the alternative approach and compare it to the FFNS$_{CJKL}$ parametrization\footnote{The FFNS$_{CJKL}$ parametrization was the first of the serie of the analysis of the structure of the photon presented by the CJKL group. The othes could be found in \cite{data_link}} \cite{kraw}.

\subsection{$FFNS_{CJKL}$ parametrization}

In \cite{kraw} the input scale is chosen small, $Q^2=0.25\ $GeV$^2$ and the input hadronic distribution are generated via Vector Meson Dominance (VDM) model

\begin{equation}
 f^{\gamma,had}(x,Q_0^2)=\sum_V \frac{4\pi\alpha}{\hat{f}^2_V}f^V(x,Q^2_0).
\end{equation}
In practice one takes into account the $\rho^0$ meson and the contributions from other mesons are accounted via a parameter $\kappa$ which is left as a free parameter
\begin{equation}
 f^{\gamma,had}(x,Q_0^2)=\kappa \frac{4\pi\alpha}{\hat{f}^2_{\rho}}f^{\rho}(x,Q^2) 
\end{equation}
where $f^2_{\rho}/4\pi\approx 2$.
The input distribution functions are parameterized as folows:
\begin{eqnarray}
 x\zeta^{\rho}(x,Q_0^2)&=&0,\\
xv^{\rho}(x,Q_0^2)&=&N_vx^{\alpha}(1-x)^{\beta},\\
xG^{\rho}(x,Q_0^2)&=&N_gx^{\alpha}(1-x)^{\beta}
\label{FFNS_param}
\end{eqnarray}
where $\zeta^{\rho}$, $v^{\rho}$ and $G^{\rho}$ stands for the sea-quark, valence-quark like and gluon distribution function of the $\rho$ meson and the following constraints are imposed
\begin{eqnarray}
 \int_0^1 2v^{\rho}(x,Q_0^2)dx=2,\\
 \int_0^1 x(2v^{\rho}(x,Q_0^2)+G^{\rho}(x,Q^2_0))dx=1.
\end{eqnarray}
The first is related to the number of valence quarks in the $\rho^0$ meson, and the second one represents the energy-momentum sum rule.
These constraints allow to express the normalization factors $N_v$ and $N_g$ as functions of $\alpha$ and $\beta$:
\begin{eqnarray}
 N_v(\alpha,\beta)&=&{\frac {\Gamma \left( \alpha+\beta+1 \right) }{\Gamma \left( \beta+1
 \right) \Gamma \left( \alpha \right) }},\\
N_g(\alpha,\beta)&=&-{\frac {\alpha\,\Gamma \left( 2+\alpha+\beta \right) -\Gamma
 \left( 2+\alpha+\beta \right) \beta-\Gamma \left( 2+\alpha+\beta
 \right) }{\Gamma \left( \beta+1 \right) \Gamma \left( 1+\alpha
 \right) \alpha+\Gamma \left( \beta+1 \right) \Gamma \left( 1+\alpha
 \right) \beta+\Gamma \left( \beta+1 \right) \Gamma \left( 1+\alpha
 \right) }}.
\end{eqnarray}
Together with the parameter $\kappa$ this leaves $\alpha$, $\beta$ and $\kappa$ as the only free parameters to be fixed in the fits to the $F_2^{\gamma}$.

The $FFNS_{CJKL}$ parametrization assumes three active massless quarks: $u$, $d$ and $s$. The hadronic part of the non-singlet and the singlet distribution functions can be described by means of \reff{FFNS_param} in the following way:

\begin{eqnarray}
 q_{NS}^{\gamma,had}(x,Q^2)&=&\frac{1}{9}v^{\gamma,had}(x,Q^2),\\
\Sigma^{\gamma,had}(x,Q^2)&=&2v^{\gamma,had}(x,Q^2)+6\zeta^{\gamma}(x,Q^2)
\end{eqnarray}

Finally, the contribution to $F_2^{\gamma}$ due to the massive c and b quarks is added. It is approximated by the Bethe-Heitler formula valid for $W^2>4m_h^2$:
\begin{equation}
 \frac{1}{x}F_{2,h}^{\gamma}(x,Q^2)=3\frac{\alpha}{\pi}e_h^4w(x,Q^2)
\end{equation}
with
\begin{eqnarray}
 w(x,Q^2)&=&\beta\left[-1+8x(1-x)-x(1-x)\frac{4m_h^2}{Q^2} \right] \nonumber \\
&&+\ln (\frac{1+\beta}{1-\beta})\left[x^2+(1-x)^2+x(1-3x)\frac{4m_h^2}{Q^2}-x^2\frac{8m_h^4}{Q^4} \right], \\
\beta&=&\sqrt{1-\frac{4m_h^2x}{(1-x)Q^2}}=\sqrt{1-\frac{4m_h^2}{W^2}}.
\end{eqnarray}

\subsection{Global fit at the LO}

In this section we briefly present the results of the QCD analysis of the structure function $F_2^{\gamma}$ in the alternative approach at the LO. We adopt the Fixed-Flavour-Number Scheme model of \cite{kraw} with the same set of the experimental data.

We have performed minimum $\chi^2$ fit using MINUIT \cite{James}. The set of 182 experimental data was used \cite{data_link}. Systematic and statistical errors were added in quadrature.

For $n_f=3$ we took the QCD scale $\Lambda^{(3)}=314$ MeV which corresponds to the value $\Lambda^{(4)}=280$ \cite{DataPartPhys} on the assumption that $\alpha_s(\mu)^{(n_f)}=\alpha_s(\mu)^{(n_f-1)}$. Masses of the heavy quarks are taken to be $m_b=1.3$ GeV and $m_c=4.3$ GeV 

\begin{table}[htb]
\begin{center}
\begin{tabular}{  | c | c | c | c | c | }
\hline
      \bf Approach & \bf $\chi^2$ &\bf $\kappa$ &\bf $\alpha$ &\bf $\beta$  \\
\hline
   Coventional  & 357  & 1.726 & 0.465 & 0.127 \\
\hline
   Alternative  & 591 & 0.876 & 1.878 & 4.788 \\
\hline
\end{tabular}
\end{center}
\caption{The $\chi^2$ for 182 points and parameters of the fit in conventional and alternative approach in the $FFNS_{CJKL}$ model.}
\label{srr1}
\end{table}

Three parameters $\alpha, \beta$ and $\kappa$ describing the input hadronic distribution function and value of $\chi^{2}$ in both approaches are presented in Tab.\ref{srr1}. In the alternative approach in the $FFNS_{CJKL}$ we have obtained higher value of $\chi^2$ then in the conventional one \cite{kraw} and notably different set of parameters of the input distribution function. In order to improve the quality of the fit one would have to look for the new parametrization.

\begin{figure}\unitlength=1mm
\begin{picture}(160,200)
\put(0,165){\epsfig{file=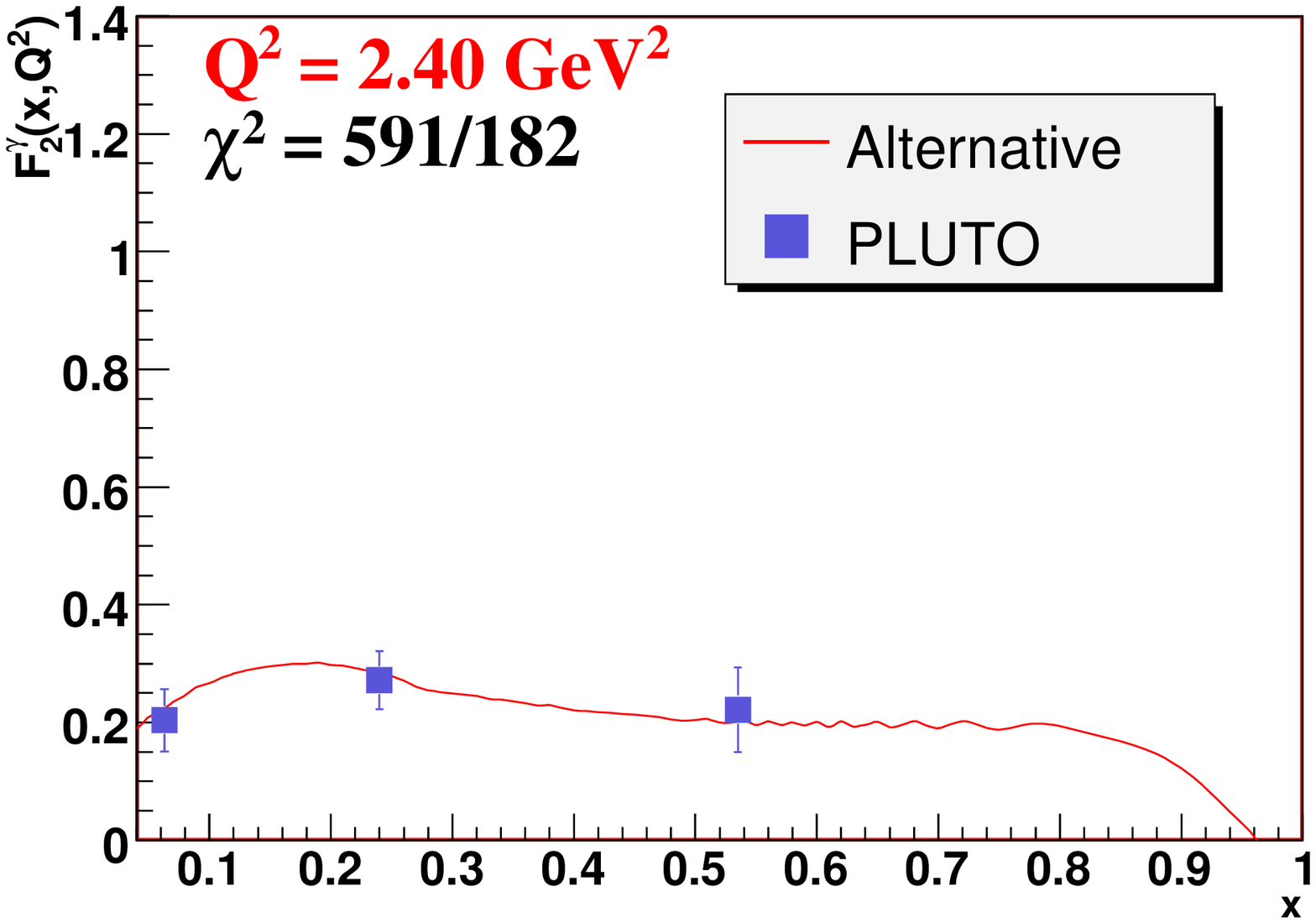,width=5cm}}
\put(50,165){\epsfig{file=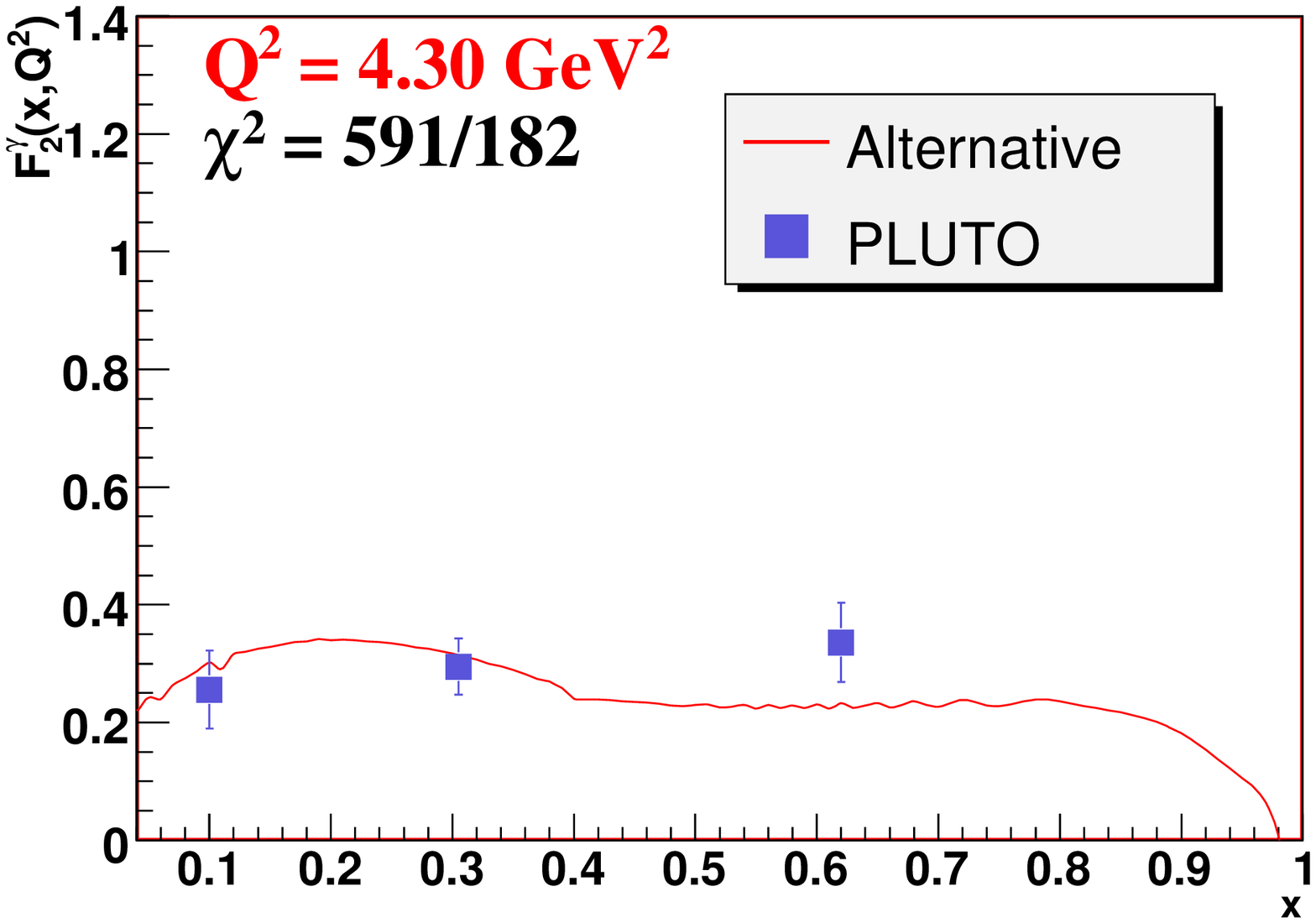,width=5cm}}
\put(100,165){\epsfig{file=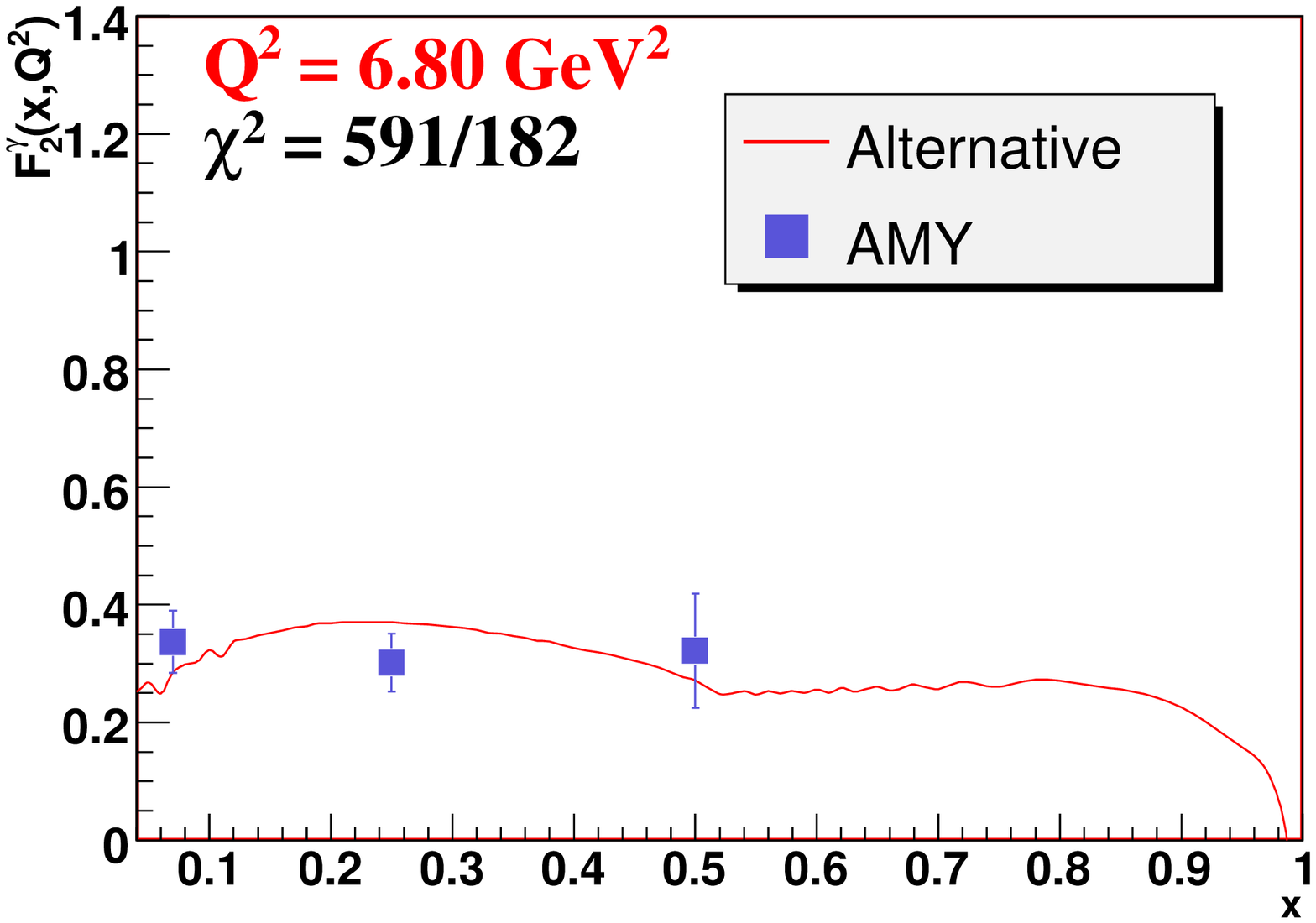,width=5cm}}
\put(0,125){\epsfig{file=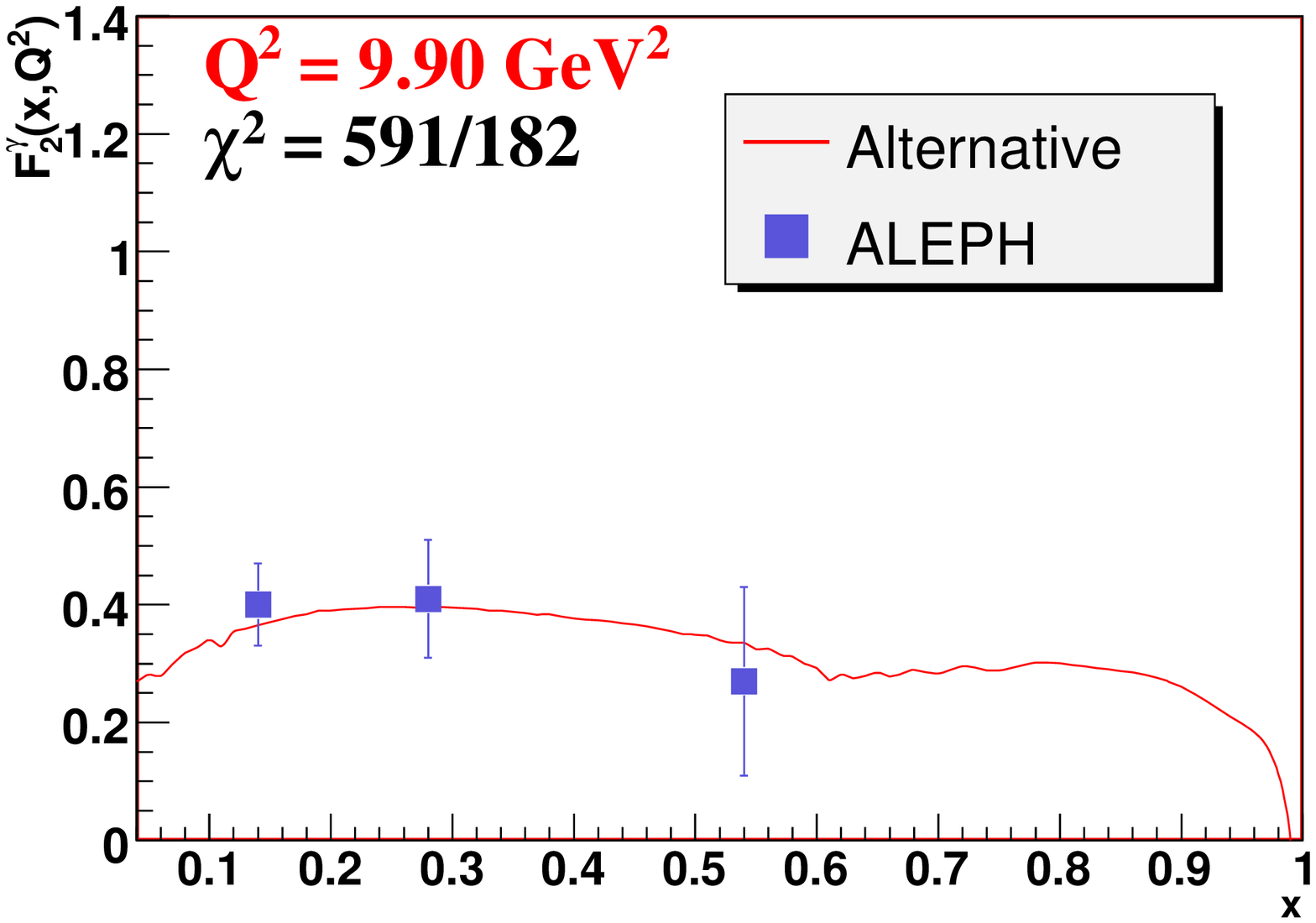,width=5cm}}
\put(50,125){\epsfig{file=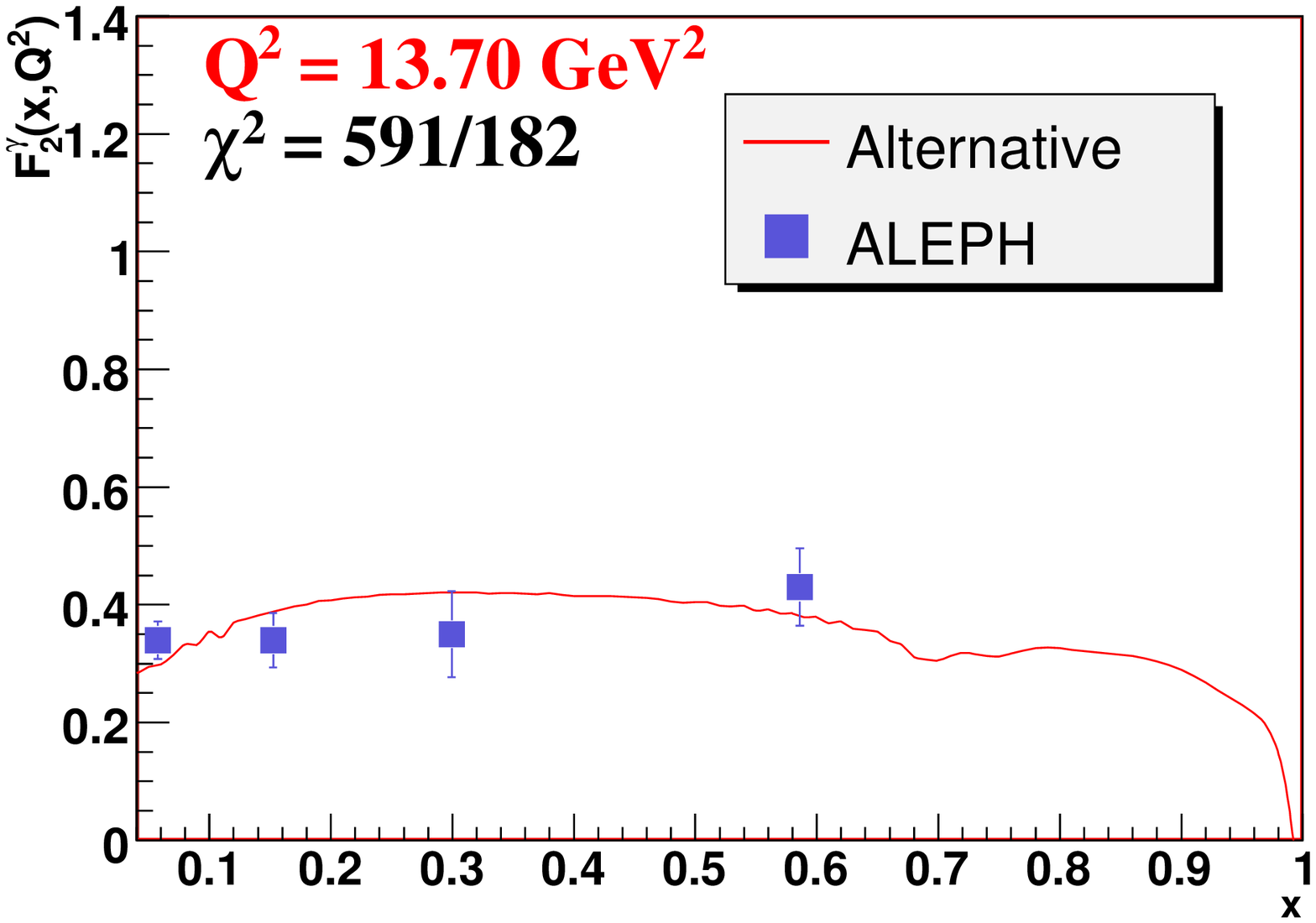,width=5cm}}
\put(100,125){\epsfig{file=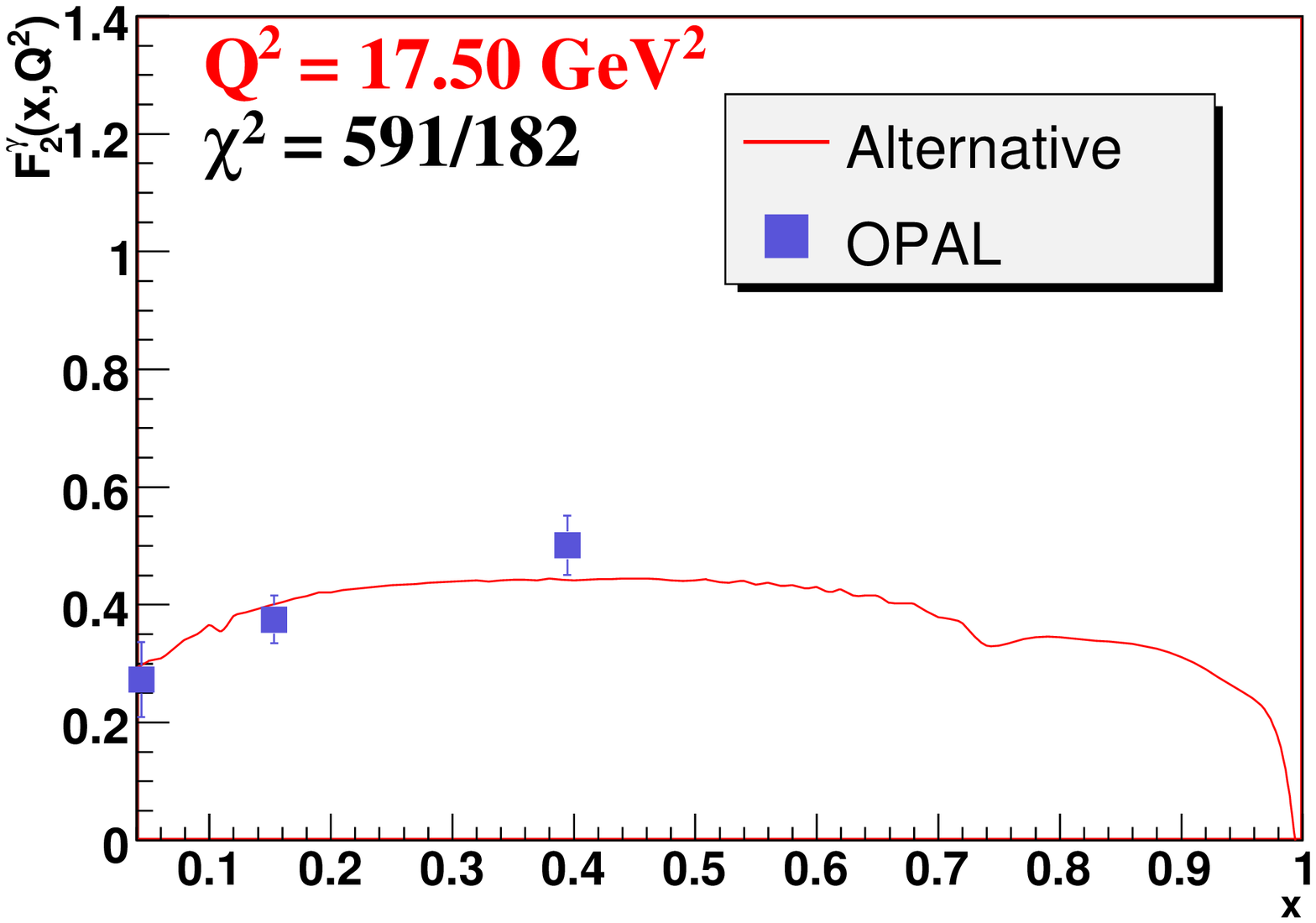,width=5cm}}
\put(0,85){\epsfig{file=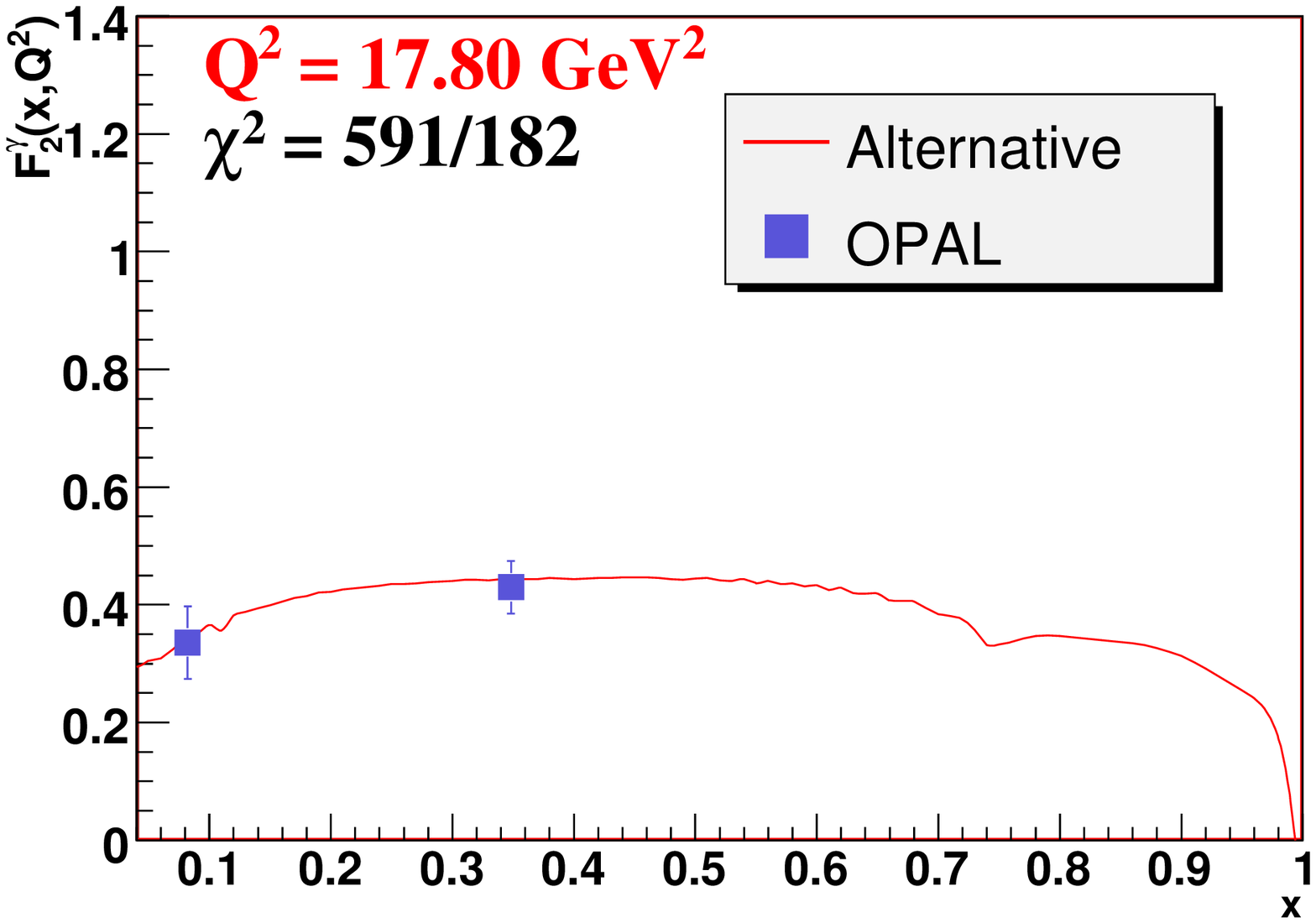,width=5cm}}
\put(50,85){\epsfig{file=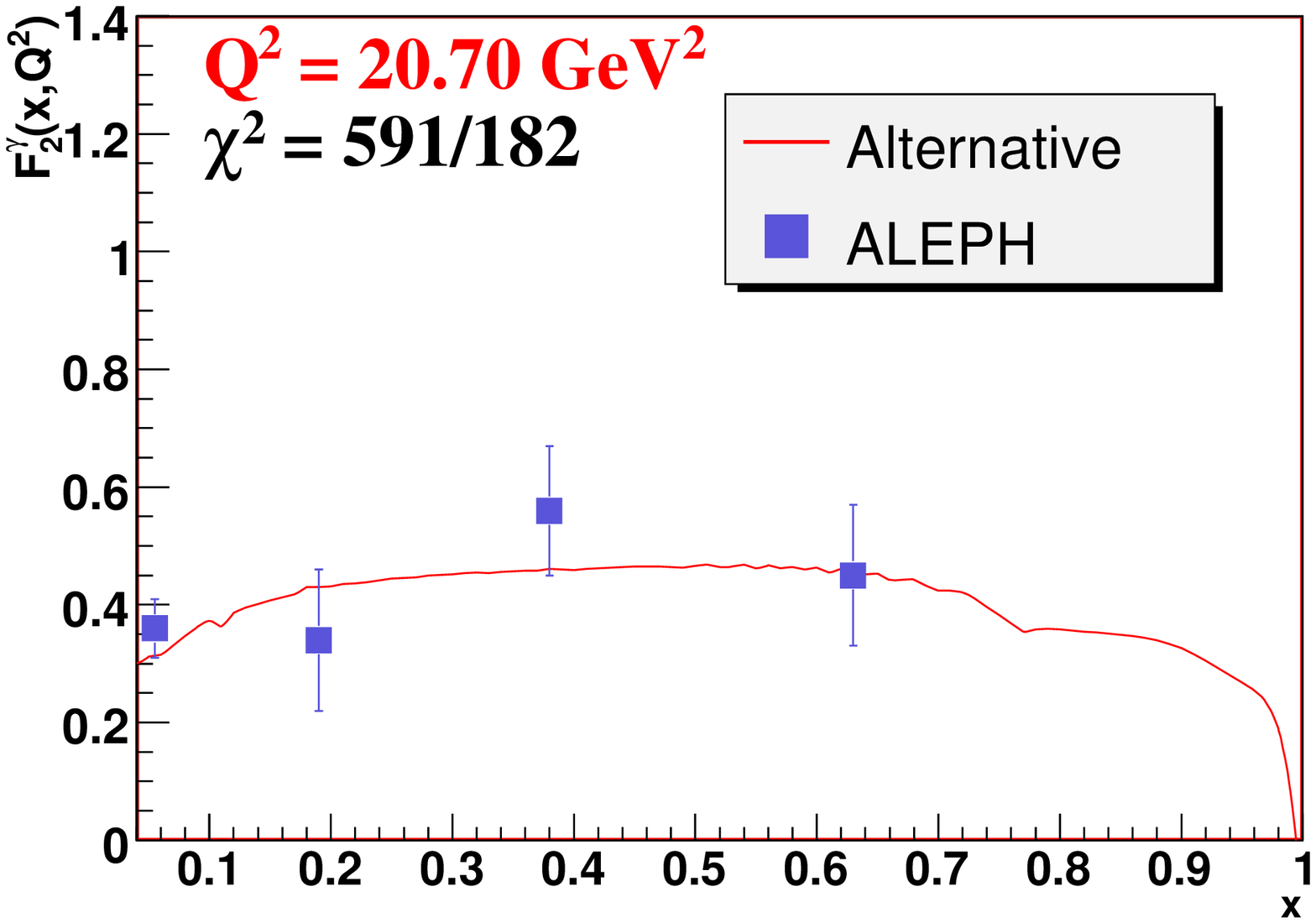,width=5cm}}
\put(100,85){\epsfig{file=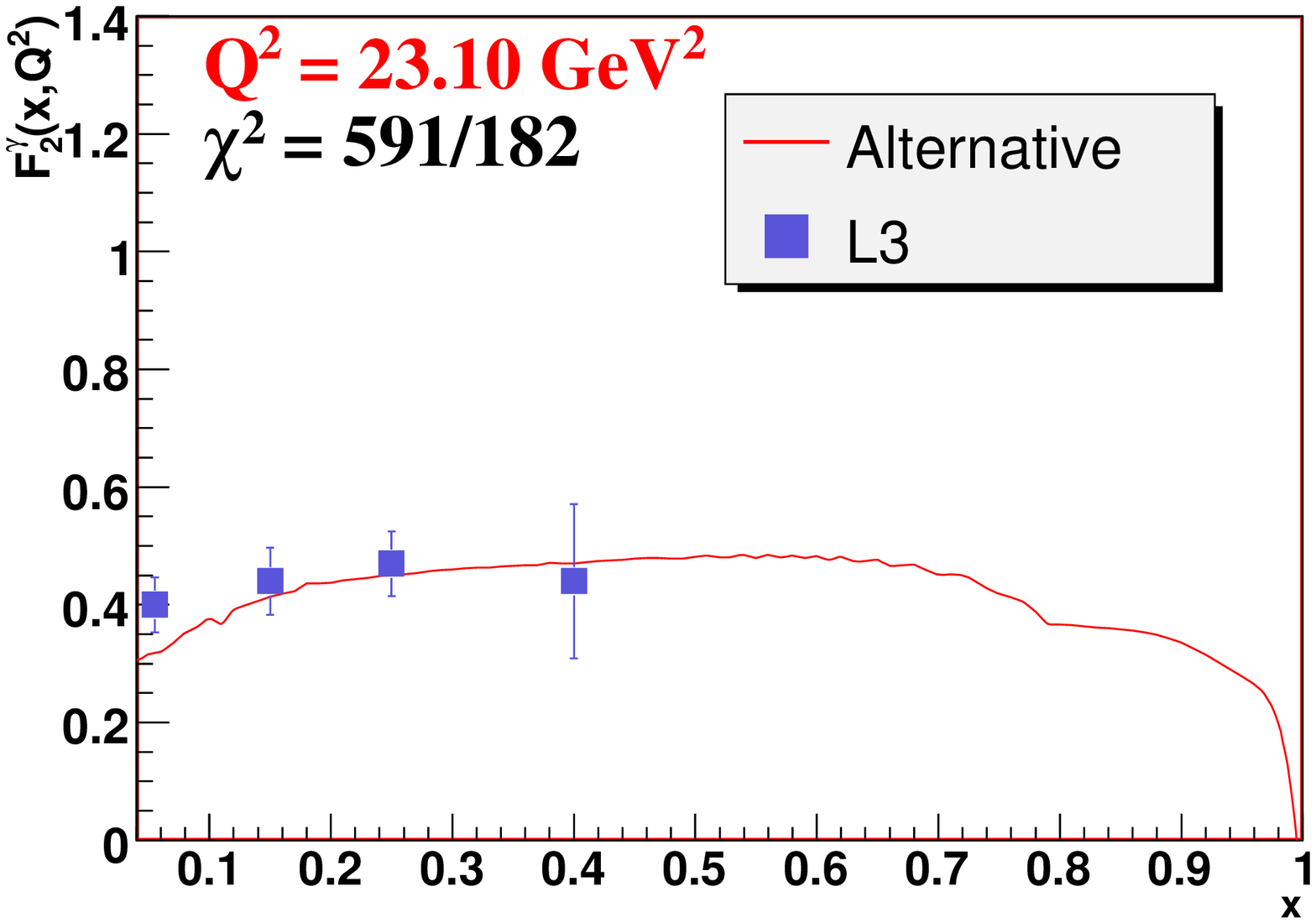,width=5cm}}
\put(0,45){\epsfig{file=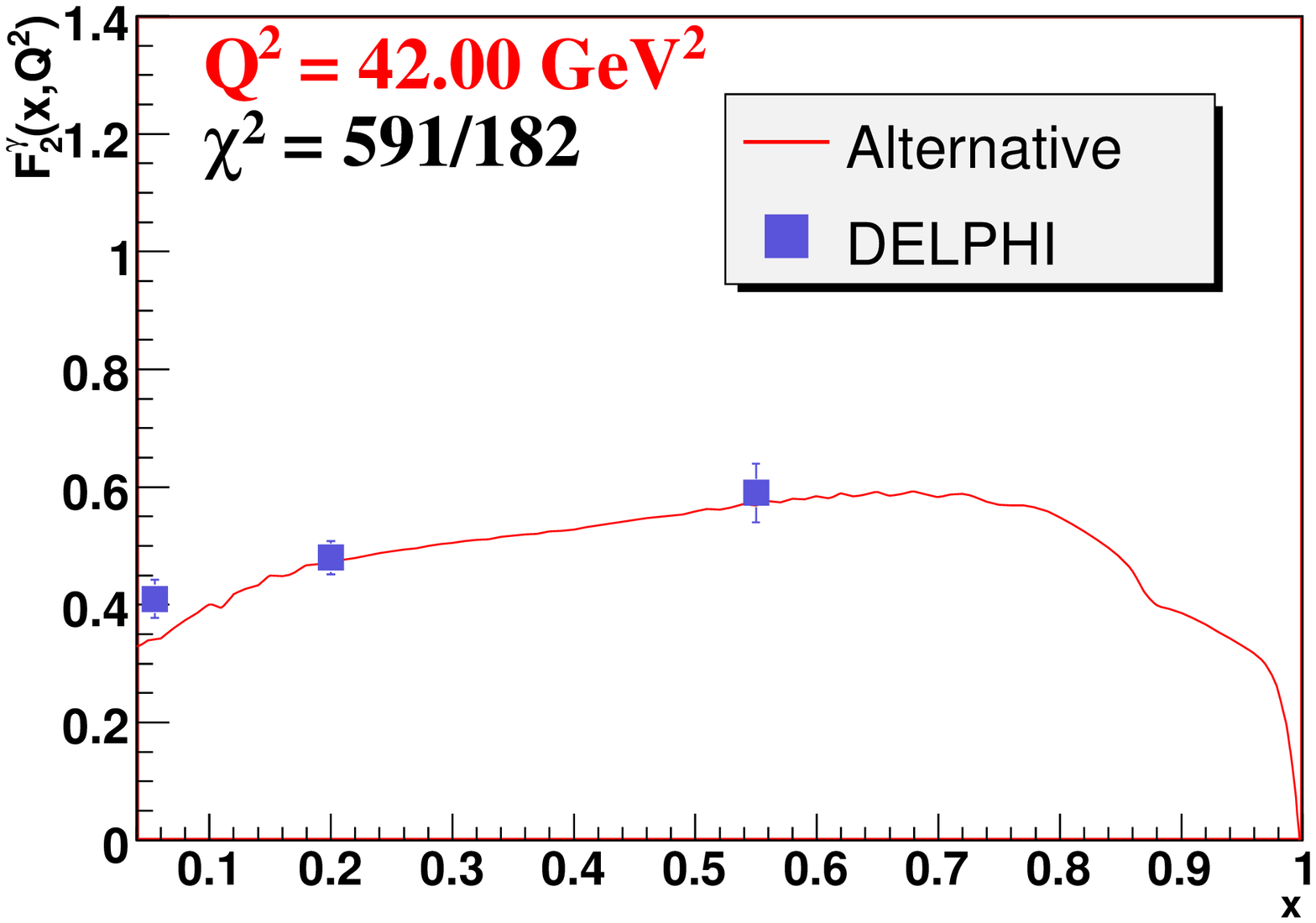,width=5cm}}
\put(50,45){\epsfig{file=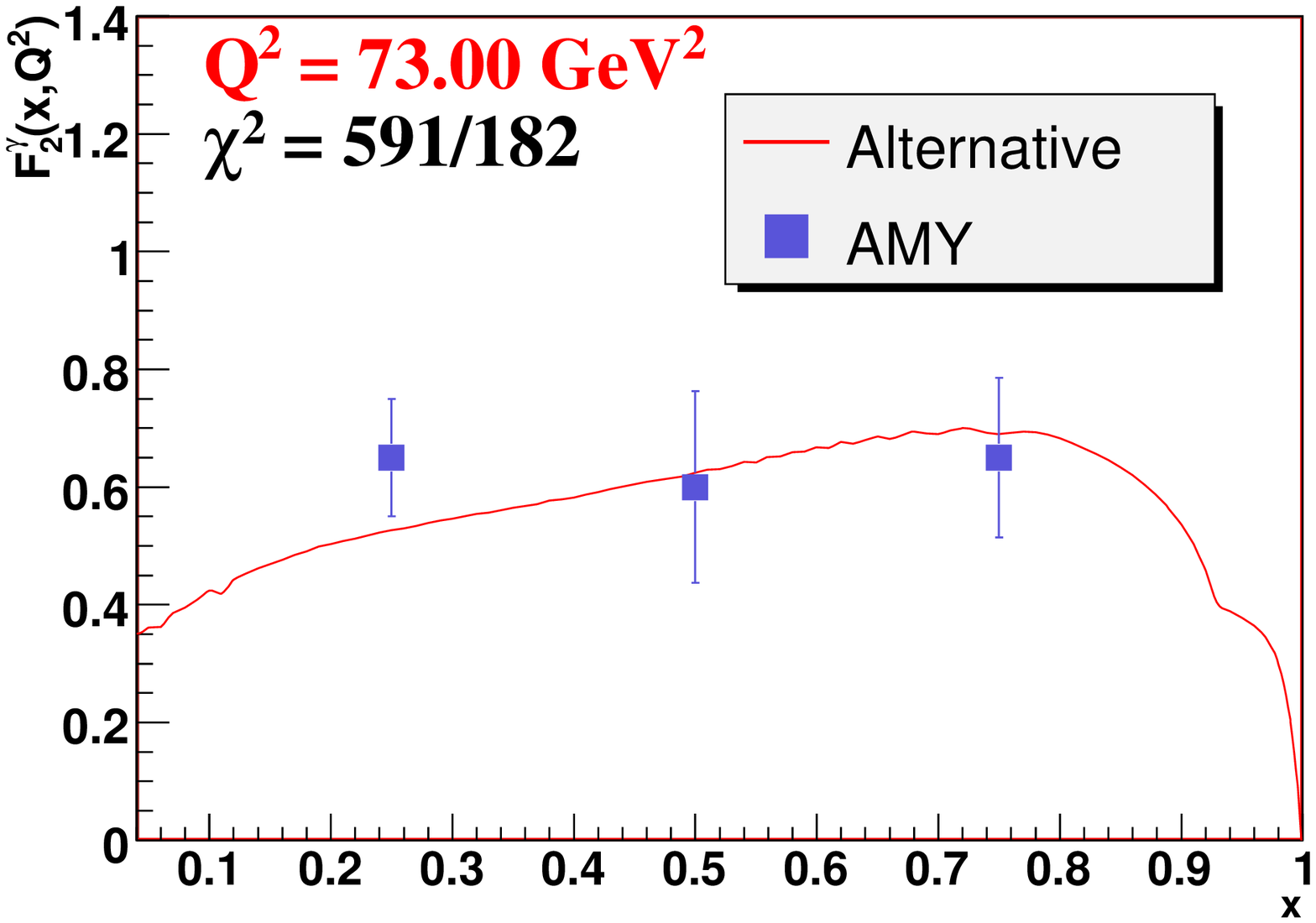,width=5cm}}
\put(100,45){\epsfig{file=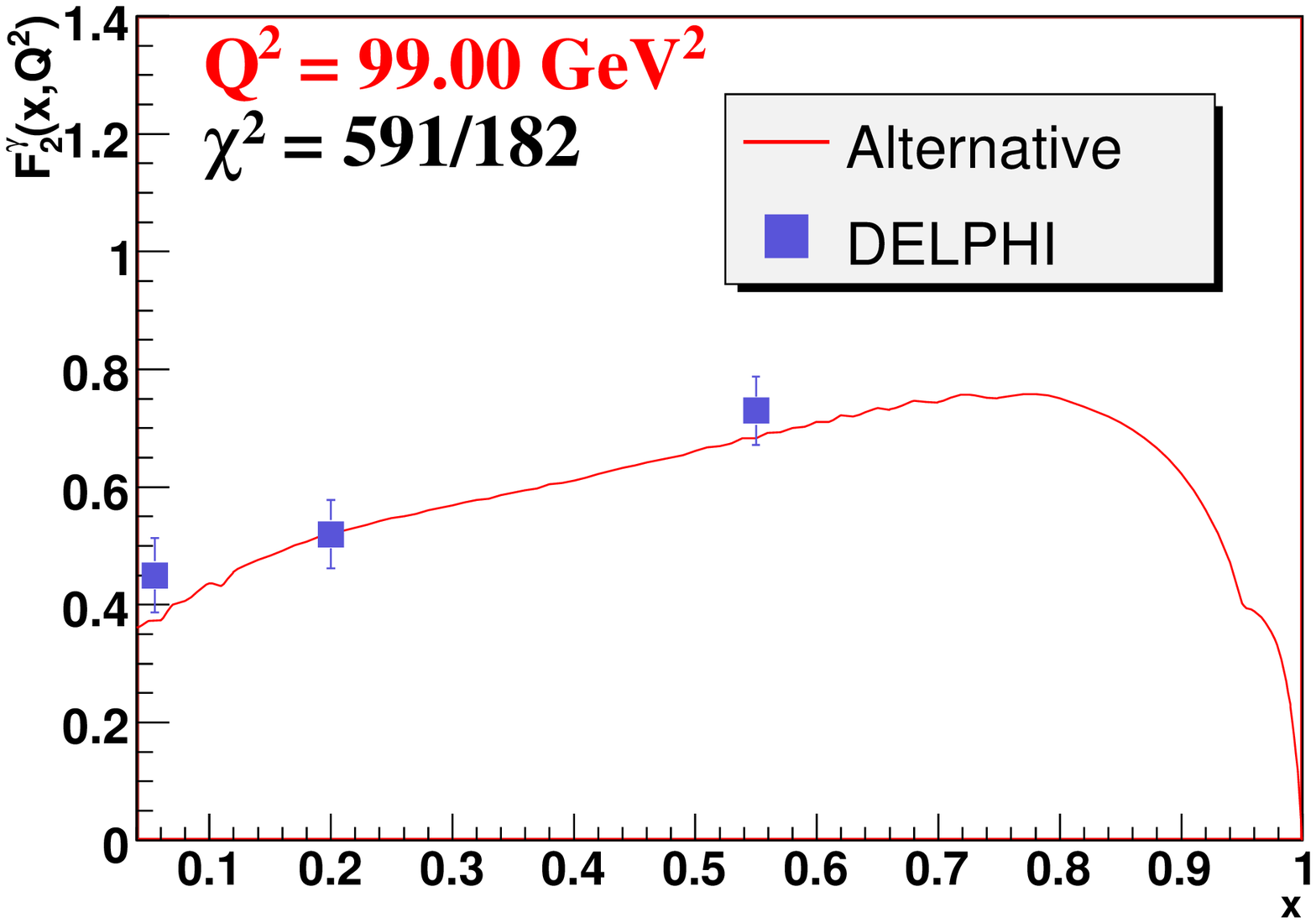,width=5cm}}
\put(0,5){\epsfig{file=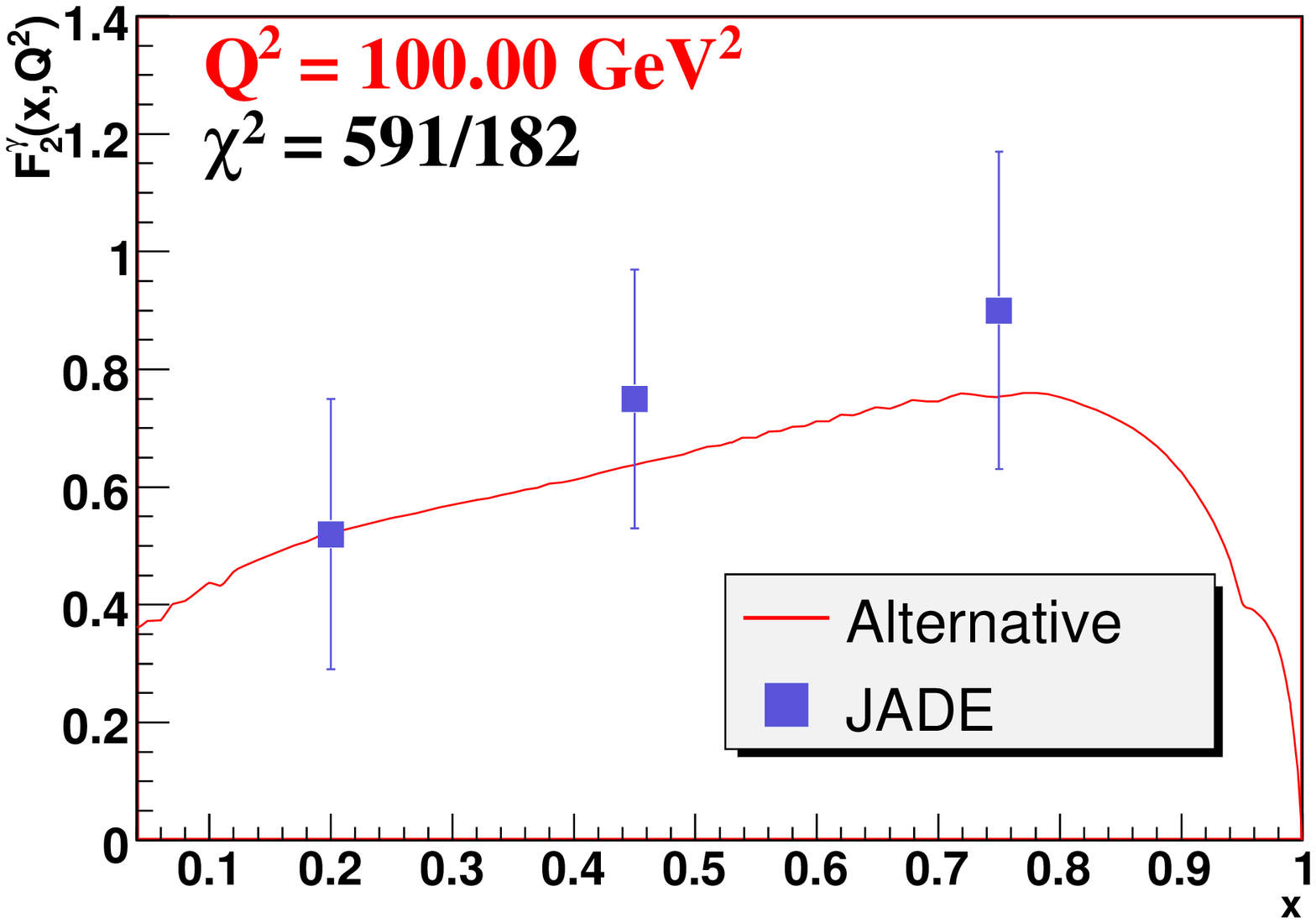,width=5cm}}
\put(50,5){\epsfig{file=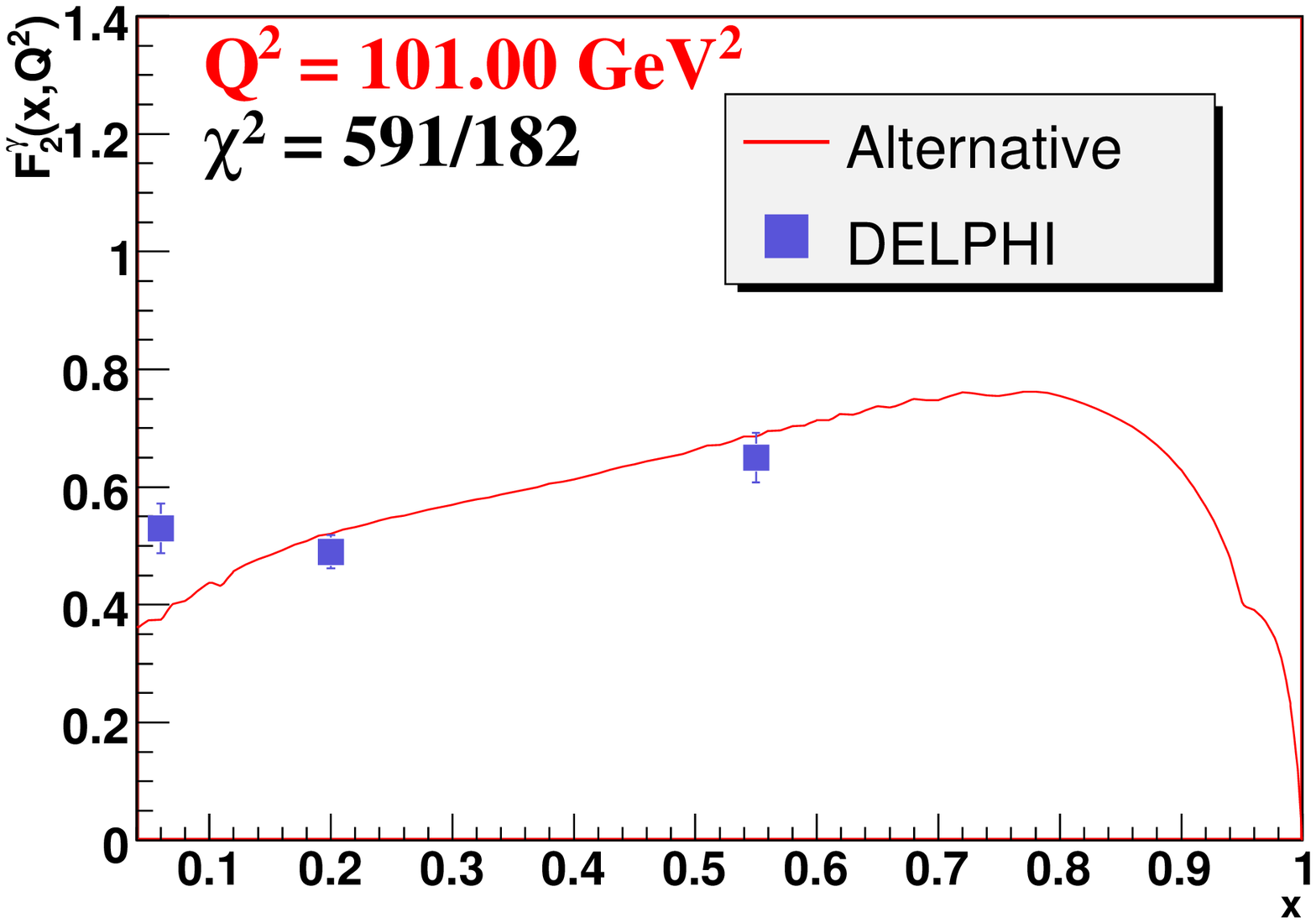,width=5cm}}
\put(100,5){\epsfig{file=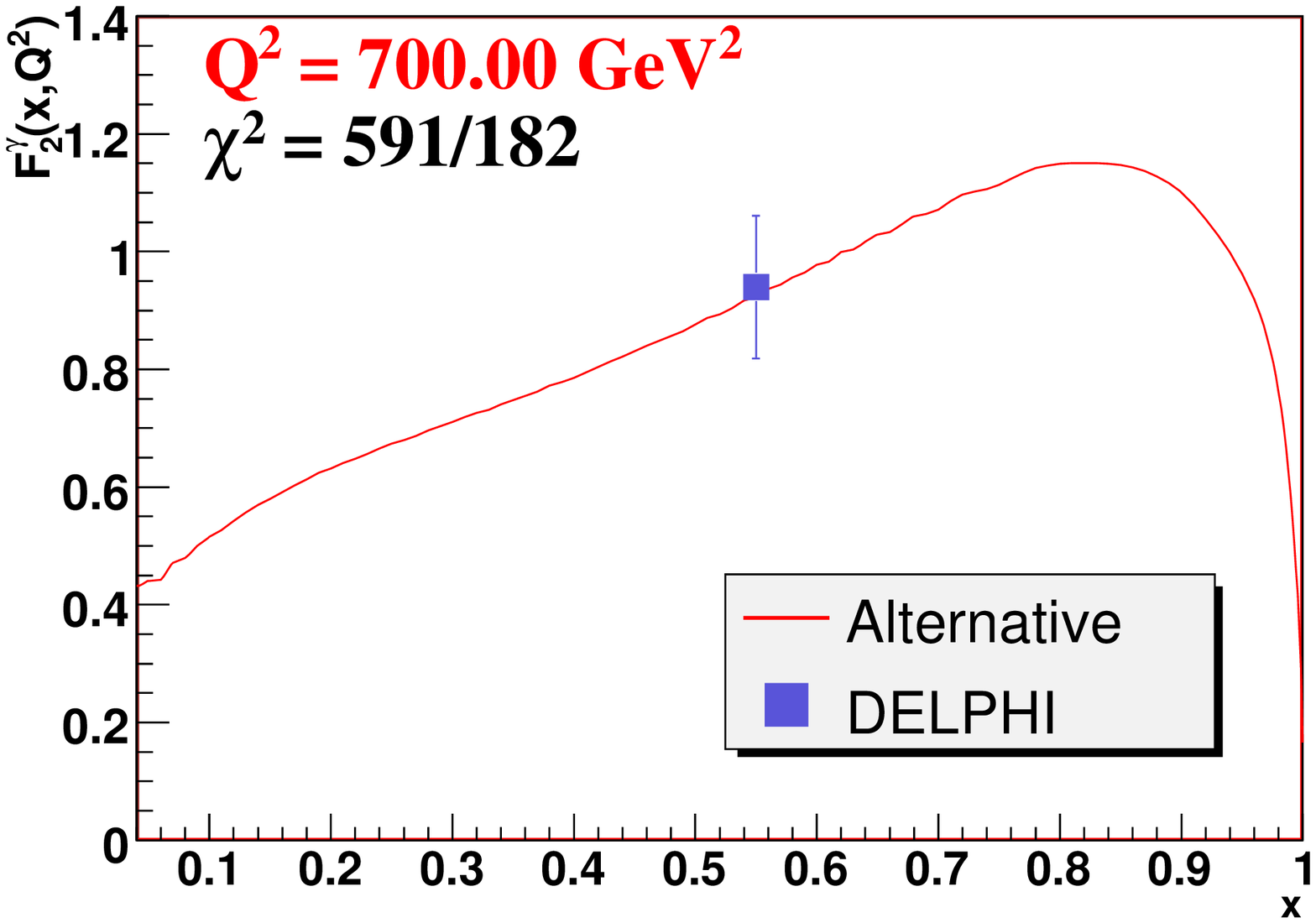,width=5cm}}
\end{picture}
\caption{Fits of experimental data of $F_2^{\gamma}$ calculated in alternative approach in the $FFNS_{CJKL}$ model.}
\label{s1}
\end{figure}

Graphical ilustration of the result of the fit is in Fig. \ref{s1}. For full set of figures, consult \cite{url}.

\newpage

\section{Summary and conclusions}
The basic idea of the alternative approach to the analysis of the photon
structure function $F_{2}^{\gamma}$, based on the separation of pure QED effects from those of genuine QCD origin, is recalled. This approach differs from the 
conventional one by reserving the meaning of the concepts of the "leading" and "next-to-leading" order to the genuine QCD effects.  

The LO and NLO approximations for $F_{2}^{\gamma}$ in both the conventional and alternative  approach are discussed in the detail and essential differences are pointed out. It is argued that the inhomogeneous splitting function $k_q^{(1)}$ should be included in the evolution equations for quark distribution functions of the photon already at the LO. Similarly, the photonic coefficient functions $C_{\gamma}^{(0)}$ (which is of pure QED origin) and $C_{\gamma}^{(1)}$ as well as quark coefficient function $C_q^{(1)}$ should be included in the LO
formula for $F_2^{\gamma}$.

At the NLO the alternative approach differs from the conventional one by the presence of the inhomogeneous splitting functions $k_q^{(2)}$ and $k_G^{(2)}$ 
in the evolution equations and by the inclusion of the terms proportional to
$C_{\gamma}^{(1)}$, $C_{\gamma}^{(2)}$, $C_q^{(2)}$ and $C_G^{(2)}$ in the
formula for $F_2^{\gamma}$. Except for $C_{\gamma}^{(2)}$, all these functions 
are known, and also $C_{\gamma}^{(2)}$ is calculable with current methods, though not quite straighforwardly \cite{Private} and there si thus hope that complete NLO analysis of $F_{2}^{\gamma}$ in the alternative approach will soon be
possible to perform.

As several quantities which in the conventional approach enter $F_{2}^{\gamma}$ 
only at the NLO, appear in the alternative approach already at the LO,  
we compare the latter with the pointlike parts of the LO as well as NLO approximations in the conventional approach. We concentrate on
the pointlike parts as this is where the two approaches differ.

As far as the quark distribution functions are concerned, the LO approximation in the alternative approach is very close to the NLO approximation of the conventional approach. This reflects the fact that the inclusion of the effect
of including the inhomogeneous splitting function $k_q^{(1)}$ is bigger than
that of the homogeneous splitting functions $P^{(1)}$. On the other hand the
LO approximation in the conventional approach is significantly lower than
LO one of the alternative approach, in particular for large $x$. At the NLO, 
however, this difference becomes negligible. The LO approximation in the alternative approach thus captures all essential
contributions to $F_{2}^{\gamma}$ and in view of the accuracy of the existing
data should thus be sufficient for its QCD analysis.

\section*{Acknowledgment}
I would like to thank Jiri Chyla for careful reading of this text and valuable comments and suggestions. This work has been supported by the projects AV0Z10100502 of the Academy of Sciences and LC527 of the Ministry of Education of the Czech Republic.

\clearpage

\newpage
  \renewcommand{\theequation}{A-\arabic{equation}}
  \setcounter{equation}{0}  
  \section*{APPENDIX A\\ Coefficient and splitting functions}  

Let us present the formulas of the splitting and coefficient functions used in our calculations. They follow the expansions \reff{k(x)}-\reff{P(x)}, \reff{expansion_Cq}-\reff{expansion_Cgamma}.

The quark coefficient functions were calculated in $\overline{MS}$ factorization scheme. We present them in x-space.

The quark coefficient function $C_q^{(1)}$ reads \cite{GRV1}:

\begin{eqnarray}
C_q^{(1)}(x,1)&=&\frac{2}{3} \left(4 \left[\frac{\ln(1-x)}{1-x}\right]_+-3 \left[\frac{1}{1-x}\right]_+-(9 + \frac{2\pi^2}{3} ) \delta(1-x ) - 2 (1 + x)\ln \frac{1-x}{x}\right. \nonumber \\
&&\left. - \frac{4\ln x}{1-x} + 6 + 4 x\right),\label{Cq1}
\end{eqnarray}
The quark coefficient function $C_q^{(2)}$ can be written approximately as \cite{VVM}:

\begin{eqnarray}
\!\!\!\!\!\!\!\!\!&&C_q^{(2)}(x,1)=   \frac{1}{4}\left(\frac{128}{9} D_3 - \frac{184}{3} D_2 - 31.1052 D_1 + 188.641 D_0 - 338.513 \delta(1-x)- 17.74 L_1^3 +  \right. \nonumber\\ 
\!\!\!\!\!\!\!\!\!&&72.24 L_1^2 - 628.8 L_1 - 181 - 806.7 x + 0.719 xL_0^4+ L_0 L_1 (37.75 L_0 - 147.1 L_1 ) - \nonumber\\
\!\!\!\!\!\!\!\!\!&&28.384 L_0 - 20.70 L_0^2 - \frac{80}{27} L_0^3+ n_f \left( \frac{16}{9} D_2 - \frac{232}{27} D_1 + 6.34888 D_0 + 46.8531 \delta(1-x) - \right. \nonumber\\
\!\!\!\!\!\!\!\!\!&& \left. \left. 1.5 L_1^2 + 24.87 L_1 - 7.8109 - 17.82 x - 12.97 x^2 - 0.185 xL_0^3 + 8.113 L_0 L_1 + \frac{16}{3} L_0 + \frac{20}{9} L_0^2 \right) \right) \nonumber
\end{eqnarray}
where 
\begin{eqnarray}
L_0=\ln x, \quad L_1=\ln (1-x), \quad D_k=\left[\frac{\ln^k(1-x)}{1-x}\right]_+.
\end{eqnarray}
The photonic coefficient function $C_{\gamma}^{(0)}$ reads:
\begin{equation}
C_{\gamma}^{(0)}(x,1)=(1-2x+2x^2)\ln \frac{1-x}{x}-1+8x(1-x)
\end{equation}
The photonic coefficient function $C_{\gamma}^{(1)}$ can be represented by the approximate formula:
\begin{eqnarray}
C_{\gamma}^{(1)}(x)=\!\!\!\!\!\!\!\!&&\frac{1}{24}\left[ (26.67 - 317.5 (1 - x))L_1^3 - 72.00L_1^2 - (1287 x^{-1} - 908.6)L_1 \right. \nonumber\\
&& + 1598L_1L_0 \left. - 13.86L^3_0 - 27.74L^2_0 - 67.33L_0 - 1576 - 2727 x + 3823 x^2 \right]
\end{eqnarray}
The photonic coefficient functions are given in Mellin moments. The leading and the next-to-leading terms are given by \cite{GRV1}:

\begin{equation}
k_{NS}^{(0)}(n)=3n_f\left(\langle e^4 \rangle-\langle e^2 \rangle^2 \right)\kappa_1, \quad k_{NS}^{(1)}(n)=3n_f\left(\langle e^4 \rangle-\langle e^2 \rangle^2 \right)\kappa_2
\end{equation}

\begin{eqnarray}
\kappa_1=2\frac{n^2+n+2}{n(n+1)(n+2)}\\
\kappa_2=\frac{4}{3}\left((S_1^2(n)-S_2(n)+\frac{5}{2})2\frac{n^2+n+2}{n(n+1)(n+2)}-\right.\nonumber\\
\left. \frac{4}{n^2}S_1(n)+\frac{11n^4+26n^3+15n^2+8n+4}{n^3(n+1)^3 (n+2)} \right)
\end{eqnarray}
The non-singlet NNLO photonic coefficient function $k_{NS}^{(2)}(n)$ was obtained by the Mellin transform of the $k_{NS}^{(2)}(x)$ presented in \cite{VVM1} :

\begin{eqnarray}
&&k_{NS}^{(2)}(n)=\nonumber\\
&&\frac{3n_f\left(\langle e^4 \rangle-\langle e^2 \rangle^2 \right)}{8}\left[{\frac {128}{27}}\,{\frac { S_1^{4} \left( n \right)+6 S_2 \left( n \right) S_1^{2} \left( n \right) +8\, S_3 \left( n \right)  S_1 \left( n \right) +3  S_2^{2} \left( n \right)  +6\, S_4 \left(n \right) }{n}}\right.\nonumber \\
&& -\frac{112}{9}\,{\frac { S_1^{3} \left( n \right) +3\, S_1 \left( n \right)  S_2 \left( n
 \right) +2\, S_3 \left( n \right) }{n}}+ 175.3\,{\frac {   S_1^{2} \left( n \right)  + S_2 \left( n \right) }{n}}- 142.3\,{\frac { S_1 \left( n \right) }{n}} \nonumber \\
&&+\frac{1353}{n}-\frac{1262}{\left( n+1 \right)} + \frac{449.2}{\left( n+2 \right)}-\frac{1445}{ \left( n+3 \right)}-\frac{325.4}{n} \left(  \zeta_3+{\frac {\zeta_2}{n}}-{\frac { S_1 \left( n \right) }{{n}^{2}}}-{\frac { S_2 \left( n \right) }{n}}- S_3 \left( n \right)  \right)  \nonumber\\
&& \nonumber\\
&& - \frac{390.8}{n}\, \left( \zeta_3+ \zeta_2\,S_1 \left( n \right) -\frac{1}{2}{\frac {  S_1^{2} \left( n \right)  +{\it S_2} \left( n \right) }{n}}- S_1 \left( n \right)  S_2
 \left( n \right) - S_3 \left( n \right)  \right)  \nonumber \\
&& -\frac{1169}{  2\,n+{n}^{2}+1 }+ 50.08\,{\frac {   S_1^{3} \left( n+1 \right) +3\, S_1 \left( n+1 \right) S_2 \left( n+1 \right) +2\, S_3 \left( n+1 \right) }{n+1}} \nonumber \\
&& -\frac{744.6}{ {n}^2}+ \frac{403.2}{n^{3}}-\frac{160}{n^{4}}+{\frac {512}{9n^{5}}}+n_f\, \left( {\frac {32}{27}}\,{\frac {   S_1^{3} \left( n \right)  +3\,{ S_1} \left( n \right)  S_2 \left( n \right) +2\, S_3 \left( n \right) }{n}} \right.\nonumber \\ 
&& + 258.142\,{\frac {  S_1^{2} \left( n \right) + S_2 \left( n \right) }{n}}+ 18.77\,{\frac { S_1 \left( n \right) }{n}}- \frac{40.035}{n}+ \frac{114.4}{ n+1}- \frac{24.86}{n+2}- \frac{53.39}{n+3}\nonumber \\
&&  +\frac{17.046}{n} \left( \zeta_3+{\frac { \zeta_2}{n}}-{\frac {S_1 \left( n \right) 
}{{n}^{2}}}-{\frac { S_2 \left( n \right) }{n}}- S_3 \left( n
 \right)  \right) + \nonumber \\
&& \frac{538.8}{n} \left( \zeta_3+\zeta_2\, S_1 \left( n \right) - {\frac { S_1^{2} \left( n
  \right) + S_2 \left( n \right) }{2n}}-S_1
 \left( n \right) S_2 \left( n \right) - S_3 \left( n
 \right)  \right) + \nonumber \\
&&\left.\left. \frac{26.63}{  2\,n+{n}^{2}+1 } - 270.0\,{\frac {  { S_1^{2}} \left( n+1 \right)  + S_2 \left( n+1 \right) }{n+1}}+ \frac{21.55}{n^2}- \frac{21.984}{n^3}+{\frac {64}{9 n^4}} \right)\right]
\end{eqnarray}
where

\begin{eqnarray}
S_1(n)&=&\gamma_E+\psi(n+1),\quad \gamma_E=0.577216, \nonumber \\
S_2(n)&=&\zeta_2-\psi^{\prime}(n+1),\quad \zeta_2=\frac{\pi^2}{6} \nonumber \\
S_3(n)&=&\zeta_3+\frac{1}{2}\psi^{\prime \prime}(n+1),\quad \zeta_3=1.202057 \nonumber
\end{eqnarray}
where $\psi(n)$, $\psi^{\prime}(n)$ and $\psi^{\prime \prime}(n)$ are the $\psi$-function and its first and second derivatives.

\newpage
  \renewcommand{\theequation}{B-\arabic{equation}}
  \setcounter{equation}{0}  
  \section*{APPENDIX B\\ Solution of evolution equations in the singlet case}  

To find solution in the case of the evolution equations for singlet distribution function $\Sigma(x,Q^2)$ and gluon distribution function $G(x,Q^2)$ is less trivial as these equations are coupled. It is useful to introduce 2x2 matrices $P^{(i)}$ and 2-component vectors $k^{(i)}$:

\begin{equation}
    P^{(i)} \equiv \left( \begin{array}{cc}
    P_{qq}^{(i)} & P_{qg}^{(i)} \\
    P_{gq}^{(i)} & P_{gg}^{(i)} \\
    \end{array} \right),\ k^{(i)} \equiv \left( \begin{array}{c}
    k_q^{(i)} \\
    k_G^{(i)} \\
    \end{array} \right)
\label{mat&vec}
\end{equation}
where $P_{qq}^{(i)}$, $P_{qg}^{(i)}$, $P_{gq}^{(i)}$, $P_{gg}^{(i)}$ and $k_{q}^{(i)}$ and $k_{G}^{(i)}$ are standard parton-parton and photon splitting functions. This allow us to write the matrix evolution equation for 2-component vector:

\begin{equation}
    q^{\gamma}_S(x,M) \equiv \left[ \begin{array}{r}
    \Sigma^\gamma(x,M) \\
    G^\gamma(x,M) \\ \end{array} \right]
\end{equation}
in the following form:

\begin{equation}
 \frac{\textup{d} q_S^{\gamma}(x,M)}{\textup{d}\ln M^2}=k(x,M)+P(M)\otimes q_S^{\gamma}(M).
\label{PDFmatice}
\end{equation}
If we keep the terms up to $P^{(1)}$ and $k^{(1)}$, the pointlike solution of \reff{PDFmatice} can be written as:

\begin{eqnarray}
&&q_{S}^{\gamma,PL}(n,M)=\frac{4\pi}{\alpha_s(M)}\left[1+\frac{\alpha_s(M)}{2\pi}\hat{U} \right]\left[1-\left(\frac{\alpha_s(M)}{\alpha_s(M_0)} \right)^{1-2P^{(0)}(n)/\beta_{0}} \right]a \nonumber\\
&&+\left[1-\left(\frac{\alpha_s(M)}{\alpha_s(M_0)} \right)^{-2P^{(0)}(n)/\beta_{0}} \right]\frac{1}{-P^{(0)}(n)}\frac{\alpha}{2\pi}\left[k^{(1)}(n)-\frac{\beta_1}{2\beta_0}k^{(0)}(n)-\hat{U}k^{(0)}(n) \right]
\end{eqnarray}
where 

\begin{equation}
a\!=\!\frac{1}{1-\frac{2}{\beta_0}P^{(0)}_{qq}(n)}\frac{\alpha}{2\pi\beta_0}k^{(0)}(n)
\end{equation}
and $\hat{U}$ stands for the 2x2 matrix:

\begin{eqnarray}
\hat{U}=-\frac{2}{\beta_0}(\hat{P_+}\hat{R}\hat{P_+} + \hat{P_-}\hat{R}\hat{P_-})+\frac{\hat{P_-}\hat{R}\hat{P_+}}{\lambda_+^n - \lambda_-^n - \frac{1}{2}\beta_0}+\frac{\hat{P_+}\hat{R}\hat{P_-}}{\lambda_-^n - \lambda_+^n - \frac{1}{2}\beta_0}
\end{eqnarray}
where

\begin{eqnarray}
\hat{R}&=&P^{(1)}-\frac{\beta_1}{2\beta_0}P^{(0)},
\end{eqnarray}
\begin{eqnarray}
\lambda_\pm&=&\frac{P_{qq}^{(0)}+P_{gg}^{(0)}\pm \sqrt{\left(P^{(0)}_{qq}-P^{(0)}_{gg} \right)^2+4P^{(0)}_{qg}P^{(0)}_{gq}}}{2},\\
\hat{P}_\pm &=& \frac{\pm(P^{(0)}-\lambda_{\mp}1)}{\lambda_+^n - \lambda_-^n}
\end{eqnarray}
Hadronic solution is given by:

\begin{eqnarray}
q^{\gamma,had}_{S}(n,M) = \left\lbrace 1+\left(\frac{\alpha_s(M)}{2 \pi}-\frac{\alpha_s(M_0)}{2 \pi}\right)\hat{U}\right\rbrace \left(\frac{\alpha_s(M)}{\alpha_s(M_0^2)}\right)^{-\frac{2P^{(0)}(n)}{\beta_0}}q^{\gamma,had}_S(n,M_0).
\end{eqnarray}

\end{document}